\documentclass[titlepage,12pt,a4paper,reqno]{article}
\usepackage{amsmath,amssymb,amsfonts,graphics,setspace,tensor}
\usepackage{slashed}
\usepackage[linktocpage=true,colorlinks,pdftex]{hyperref}
\usepackage{xcolor}
\colorlet{linkequation}{blue}
\usepackage{bm}
\usepackage{doi}
\usepackage{hyperref}
\hypersetup{colorlinks, citecolor=violet, filecolor=black, linkcolor=black, urlcolor=blue}
\usepackage{chngcntr}
\newcommand*{\refeq}[1]{%
  \begingroup
    \hypersetup{
      linkcolor=linkequation,
      linkbordercolor=linkequation,
    }%
    \ref{#1}%
  \endgroup
}
\allowdisplaybreaks

\begin{document}
%\numberwithin{equation}{section} 

%-------- TITLE PAGE ----------------------------------

\begin{titlepage}

\centerline{
\Large \bf Nonlinear dynamics of a charged particle \bf} 
\medskip
\centerline{\Large \bf in a strong non-null knot wave background \bf}
%\medskip
%\centerline{\large \bf  \bf}
\vskip 1cm
\centerline{\bf Adina V.~Cri\c{s}an$^{*}$ and Ion V.~Vancea$^{\dagger}$}
\vskip 0.5cm
\centerline{\sl $^{*}$Department of Mechanical Systems Engineering}
\centerline{\sl Technical University of Cluj-Napoca}
\centerline{\sl 103 – 105 Muncii Bld., Cluj-Napoca, Romania} 
\centerline{\sl $^{\dagger}$Group of Theoretical Physics and Mathematical Physics,
Department of Physics}
\centerline{\sl Federal Rural University of Rio de Janeiro}
\centerline{\sl Cx. Postal 23851, BR 465 Km 7, 23890-000 Serop\'{e}dica - RJ,
Brazil}
\centerline{
\texttt{\small adina.crisan@mep.utcluj.ro;
ionvancea@ufrrj.br
} 
}
\vspace{0.5cm}

\centerline{26 June 2020}

\vskip 1.4cm
\centerline{\large\bf Abstract}

In this paper, we study the dynamics of the charged particle interacting with the non-null electromagnetic knot wave background. We analyse the classical system in the Hamilton-Jacobi formalism and find the action, the linear momentum and the trajectory of the particle. Also, we calculate the effective mass and the emitted radiation along the knot wave. Next, we quantize the system in the classical strong knot wave background by using the strong-field QED canonical formalism.
We explicitly construct the Furry picture and calculate the Volkov solutions of the Dirac equation. As an application, we discuss the one-photon Compton effect where we determine the general form of the $S$-matrix. Also, we discuss in details the first partial amplitudes in the transition matrix in two simple backgrounds and show that there is a pair of states for which these amplitudes are identical.

\vskip 0.7cm 

\noindent

\end{titlepage}

%------------- END TITLE PAGE -----------------------------

\section{Introduction}

Very recently, laser beams with knotted polarization singularities have been produced experimentally with torus and figure-eight knot topologies \cite{Larocque:2018}. Previously, tori knotted phase singularities in lasers had been observed in some experiments \cite{Leach:2004,Dennis:2010} following earlier theoretical studies in the field theory \cite{Nye:1974,Fadeev:1997,Berry:2001a,Berry:2001b,Flossmann:2008,Nye:2011,Dennis:2017,Bode:2017}. These exciting results obtained with the polarization and phase knots open up the possibility of generating other electromagnetic knot fields in laser experiments. 

An interesting type of knot solutions of Maxwell's equations with Hopf topology were discovered by Trautman and Ra\~{n}ada some time ago \cite{Trautman:1977im,Ranada:1989wc,Ranada:1990} and have enjoyed a constant attention since. Several studies have focused on the existence and applications of these topological electromagnetic fields to atmospheric physics \cite{Ranada:1996}, liquid crystals \cite{Irvine:2014}, plasma physics \cite{Smiet:2015}, optics \cite{Ren:2008zzf,deKlerk:2017qvq},  superconductivity \cite{Trueba:2008sc} and fluid physics \cite{Alves:2017ggb,Alves:2017zjt}. The mathematical properties of the Hopf-Ra\~{n}ada solutions and of the linked and knotted fields were studied in \cite{Ranada:1992hw,Ranada:1995,Ranada:1997,Irvine:2008,Irvine:2010}. 
The creation and dynamics of knotted electromagnetic fields were analysed in \cite{Kleckner:2013,Arrayas:2011ci,Arrayas:2017wtr,Ranada:2017ore}
and their topological quantization was discussed in \cite{Ranada:1998vp,Ranada:2003,Ranada:2006tq,Arrayas:2012eja}.
The Hopf-Ra\~{n}ada solutions have been generalized to the null and non-null torus geometries in \cite{Arrayas:2011ia,Kedia:2013bw,Hoyos:2015bxa}. More recently, new generalizations of the Hopf-Ra\~{n}ada fields have been shown to exist in the non-linear electrodynamics, fluid physics \cite{Alves:2017ggb,Alves:2017zjt,Goulart:2016orx} as well as in gravitating systems \cite{Kopinski:2017nvp,Vancea:2017tmx,Silva:2018ule,Alves:2018wku,Lechtenfeld:2017tif,Kumar:2020xjr}. (For recent reviews of the knot solutions and their applications see \cite{Arrayas:2017sfq,Vancea:2019zdl,Vancea:2019yjt} and the references therein.) 

An important problem that concerns the Ra\~{n}ada solutions and their generalizations is the understanding of the physical properties of the monochromatic knots and their interaction with matter, in particular with charged particles. Some preliminary results on the study of the motion of a classical charged particle in an electromagnetic knot were presented in \cite{Arrayas:2010xi} and a monochromatic knot solution was given recently in \cite{Cameron:2018}. The Ra\~{n}ada solutions admit a Fourier decomposition in terms of knot waves that correspond to oscillating modes  
of the electromagnetic field. These waves differ from the plane waves in that the polarization vectors carry information about the topological structure of the knot. For the classical particle, the knot structure induces a parametrized family of Finsler geometries in which one can identify pairs of geometries that are dual under the electric-magnetic duality \cite{Crisan:2020krc,CrisanA:2020}. Since the Finsler geometry is intrinsically related to the particle dynamics, it is important to see how the topological structure of a single knot wave determines the dynamics and the observables. The same problem can be posed in the quantum case. 

The purpose of this paper is to analyse the dynamics of a charged particle in a single knot wave at classical as well as quantum level. To this end, we will consider a single real wave of the non-null generalization of the Ra\~{n}ada solution discussed in \cite{Arrayas:2011ia,Arrayas:2014,Arrayas:2018}. Our motivation for this particular choice of non-null waves is two fold. On one hand, by particularizing the knot numbers to a diagonal subset of values, we obtain a real wave which is a null field and thus many conclusions obtained in this case can be applied to null knot waves as well. On the other hand, there is a practical interest in the real fields from the point of view of the experimental applications, e. g. in laser physics where the lasers can generate real monochromatic waves in a good approximation. 

There are several novel results that are presented in this paper. Firstly, the classical analysis of the charged particles in knot waves is very important to understanding the systems with knot solutions. There has been an intense activity recently in studying particular knot solutions, but no work has been reported on the electrodynamics in knot waves so far. Secondly, we give here for the first time the strong-field QED formulation of a spin half particle in a strong knot wave. 
For frequencies above a certain critical value estimated in this paper and which depends on the charge and mass of particle, the knot background is a strong classical field. These conditions, which are interesting from both theoretical and experimental point of view, allow us to treat the system in the framework of the strong-field QED. 
Although the strong-field QED method is well known, its application to the knot fields presented here gives new results from which the most important one is the derivation of the Volkov functions. The analytic form of these solutions of the Dirac equation is known only for a few classical backgrounds, among which are the plane waves. Here, we give a new Volkov function for the particular type of knot waves that contain topological information about the fields. That is an addition to the body of the analytically Volkov solutions known in the literature. And lastly, the results from this work could be of interest to the laser theory and other strong-field systems. We note that, as we show in the example of the Compton effect, there are at least some partial amplitudes in the $S$-matrix that take the same values for different states of the system and different knot waves, which indicates that the $S$-matrix could have some symmetry structure related to the topological properties of the knot wave background.  

The paper is organized as follows. In Section 2, we will construct the real electromagnetic potential of a monochromatic real knot wave of the non-null solution from \cite{Arrayas:2011ia,Arrayas:2014} and determine some useful properties of it. In Section 3, we analyse the dynamics of a classical charged particle in the knot mode background in the Hamilton-Jacobi formalism. Here we make two natural assumptions: the first one is that the charge is carried by a test particle which does not perturb the knot wave. We discuss this point in more details in the next section, where we establish a value for the critical frequencies for which the background is strong and thus stable. The second assumption is that the particle couples minimally with the knot wave. Indeed, since the knot wave is just an electromagnetic wave with additional information in the helicity vectors, the coupling can be considered a particular case of the general gauge coupling principle. The same conditions were used to discuss the classical dynamics in a general knot background in \cite{Arrayas:2010xi}. By using these hypothesis, we calculate the action, the linear momentum and the trajectory of the particle and determine the components of its trajectory in the longitudinal and transversal directions. Also, we calculate the effective mass of the particle in the knot wave and give the formula for the energy scattered per unit frequency. From it, we calculate the emitted radiation along the knot wave. In Section 4, we quantize the spin half particle in a strong knot wave background. Firstly, we show that the knot wave interacts strongly with the particle for frequencies above a critical value which we estimate. Next, we quantize the system by using the canonical method of the strong-field QED formalism. Here, we give the Furry picture, the equations of motion and we construct the Fock space. Also, we solve the Dirac equation in the presence of the knot wave and obtain the Volkov solutions. From that, the $S$-matrix formalism can be constructed by applying the standard QED method. We exemplify the $S$-matrix formalism by discussing the non-linear one-photon Compton effect in Section 5. Here, we determine the general expression of the transition matrix
and calculate explicitly the probabilities of the first partial transition amplitudes. We analyse in detail the case of the simplest non-trivial knot waves with the knot numbers $(m=0, s= 1)$ and $(m=1,s =0)$ for which we show that there are at least two equivalent quantum states of the system in the sense that the probabilities of these partial processes and the effective masses for the charged particles are identical. In the last section, we draw our conclusions and outline some future problems. For completeness, some details on the Fourier decomposition of non-null knots are presented in the Appendix A and the basic relations used to calculate the form of a partial amplitude in terms of the knot parameters are collected in the Appendix B. In the Appendix C, we summarize some basic properties of the Bessel functions. Throughout this paper, we use the natural units with $\hbar = c = 1$ and the mostly plus convention for the Minkowski metric.

\section{Real knot wave background}

In this section, we present the non-null knot solution of Maxwell's equations in vacuum obtained in \cite{Arrayas:2011ia,Arrayas:2014}. We focus on the real wave component of a single knot and establish some useful relations for it following the Fourier analysis from \cite{Arrayas:2018}. For completeness, some filling in details are presented in the Appendix A. As mentioned in the introduction, the interest in the real waves is motivated by the need of understanding the electrodynamics in knot waves as well as by possible applications to lasers.

Consider the Maxwell equations in the absence of sources  
\begin{equation}
\partial_{\mu} \mathsf{F}^{\mu \nu} = 0
\, ,
\qquad
\frac{1}{2} \epsilon^{\mu \nu \rho \sigma} 
\partial_{\nu} \mathsf{F}_{\rho \sigma} = 0
\, .
\label{Maxwell-equations-vacuum}
\end{equation}
The electromagnetic field and its dual can be expressed in terms of two potential four vectors $\mathsf{A}^{\mu} = (\mathsf{A}^{0}, \mathbf{A})$ and $\mathsf{C}^{\mu} = ( \mathsf{C}^{0} , \mathbf{C} )$ which are dependent on each other due to the electric-magnetic duality of the equations (\refeq{Maxwell-equations-vacuum}) above. Let $(m, n, l, s)$ be a quadruplet of arbitrary positive integers. Then, as shown in \cite{Arrayas:2011ia,Arrayas:2014}, the equations (\refeq{Maxwell-equations-vacuum}) admit a solution that represents the evolution of an initial electromagnetic field with knotted electric and magnetic field lines. In the Lorentz gauge $\partial \cdot \mathsf{A} = \partial \cdot \mathsf{C} = 0$ and in a reference frame where $\mathsf{A}^{0} = \mathsf{C}^{0} = 0$, the initial knot field has the following form
\cite{Arrayas:2014}
\begin{align}
\mathbf{E} (x^{0} = 0, \mathbf{x}) & = 
\frac{4}
{
\pi 
\left( 
1 + \vert \mathbf{x} \vert^2 
\right)^3
}
\begin{bmatrix}
	l\left[ 
		(x^{1})^2 - (x^2)^2 - (x^3)^2 + 1 	
	\right] 
	\\
	2\left(
		l x^{1} x^{2} - s x^{3}	
	\right)
	\\
	2\left(
		l x^{1} x^{3} + s x^{2}	
	\right)
\end{bmatrix}
\, ,
\label{E-non-null-knot}
\\
\mathbf{B} (x^{0} = 0, \mathbf{x}) & = 
\frac{4}
{
\pi 
\left( 
1 + \vert \mathbf{x} \vert^2 
\right)^3
}
\begin{bmatrix}
	2\left( 
		m x^{2} - n x^{1} x^{3}  	
	\right) 
	\\
	- 2\left(
		m x^{1} - n x^{2} x^{3}	
	\right)
	\\
	n \left[
		(x^{1})^2 + (x^{2})^2 - (x^{3})^2 + 1
	\right]
\end{bmatrix}
\, ,
\label{B-non-null-knot}
\end{align}
where $\{x^{0},x^{1}, x^{2}, x^{3} \}$ are the dimensionless coordinates in the characteristic length units for the knot field. Since the numbers 
$m$, $n$, $l$ and $s$ are arbitrary, the fields $\mathbf{E} $ and $\mathbf{B}$ are in general non-null. Since in the rest of the paper we are going to use only the $\mathsf{A}^{\mu}$ potential, we will discuss only its real monochromatic modes. The discussion of $\mathsf{C}^{\mu}$ can be found in the Appendix A. The real potential of a single knot mode $k^{\mu} = ( \omega, \mathbf{k} )$ can be written as 
\begin{equation}
\mathsf{A}^{\mu} (x, k) =
\sqrt{\frac{2}{ \omega }}
\Re{
\left[
\epsilon^{\mu}_{+} \left( k \right)
e^{-i k \cdot x}
\right]
}
\, .
\label{A-knot-mode-trig}
\end{equation}
The frequency and the wave vector satisfy the standard dispersion relations in vacuum $\omega =  \vert \mathbf{k} \vert$ and the vector $\epsilon^{\mu}_{+} \left( k \right)$ is defined in the Appendix A. One can write the potential from the relation (\refeq{A-knot-mode-trig}) above as follows
\begin{equation}
\mathsf{A}^{\mu} (x, k ) 
= \mathsf{A}^{\mu}_{1} (x, k ) + \mathsf{A}^{\mu}_{2} (x, k )
= \mathsf{A}_{knot}	\left[
			\varepsilon^{\mu}_{1,knot} (k) \cos(k \cdot x)
			-
			\varepsilon^{\mu}_{2,knot} (k) \sin (k \cdot x)
			\right]
\, ,		
\label{knot-mode-potential}
\end{equation}
with the following notations
\begin{equation}
\mathsf{A}_{knot} = \frac{e^{-\omega}}{\sqrt{2 \pi} \omega^2} 
\, ,
\qquad
\varepsilon^{\mu}_{1, knot} (k) = \left( 0 , \boldsymbol{\varepsilon}_{1, knot} (\mathbf{k}) \right)
\, , 
\qquad
\varepsilon^{\mu}_{2, knot} (k) = \left( 0 , \boldsymbol{\varepsilon}_{2. knot} (\mathbf{k}) \right)
\, . 
\label{knot-amplitude-epsilons}
\end{equation}
By using the relations given in the Appendix A and after a short algebra, we can see that the three dimensional vectors 
$\boldsymbol{\varepsilon}_{1, knot} \left( k \right)$ 
and 
$\boldsymbol{\varepsilon}_{2,knot} \left( k \right)$ 
have the following form
\begin{equation}
\boldsymbol{\varepsilon}_{1, knot} \left( k \right) =
\begin{bmatrix}
	m k_1 k_3
	\\
	m k_2 k_3 + s \omega k_3
	\\
	- m \left( k^{2}_{1} + k^{2}_{2} \right) - s \omega k_2 
\end{bmatrix}
\, ,
\qquad
\boldsymbol{\varepsilon}_{2,knot} \left( k \right) =
\begin{bmatrix}
	-n \omega k_2  -l \left( k^{2}_{2} + k^{2}_{3} \right) 
	\\
	n \omega k_1 +l k_1 k_2 
	\\
	l k_2 k_3
\end{bmatrix}
\, .
\label{epsilon-3-vectors}
\end{equation}
For general values of the parameters $m$, $n$, $l$ and $s$, the sine and cosine waves from (\refeq{knot-mode-potential}) are not orthogonal on each other and more complicate non-linearities appear in the system. In the rest of the paper, we are going to consider the case with $l = s$ and $n = m$. Then one can easily show that the  four dimensional vectors satisfy the following relations
\begin{align}
& \varepsilon_{1, knot} (k) \cdot \varepsilon_{2, knot} (k) = 0
\, ,
\label{scalar-product-epsilons}
\\
& \vert \boldsymbol{\varepsilon}_{1, knot} (\mathbf{k}) \vert^{2}
= 
\vert \boldsymbol{\varepsilon}_{2, knot} (\mathbf{k}) \vert^{2}
=
\omega^{2}
\left[
	m^{2} \left( k^{2}_{1} + k^{2}_{2} \right)
	+
	s^{2} \left( k^{2}_{2} + k^{2}_{3} \right)
	+
	2 s m \omega k_{2}
\right]
\, .
\label{knot-amplitude-epsilons-properties}
\end{align}
Also, since in the Lorentz gauge $k \cdot \mathsf{A} (x,k) = 0$, 
it follows that each component of $\mathsf{A}^{\mu} (x, k )$ is 
orthogonal on the wave four vector
\begin{equation}
k \cdot \mathsf{A}_{1}(x, k ) = k \cdot \mathsf{A}^{\mu}_{2} (x, k )
= 0
\, .
\label{Lorentz-gauge-k-A1-A2}
\end{equation}
We note that the real knot wave does not depend on the numbers $n$ and $l$. Moreover, it follows from the above relations that $\{ \boldsymbol{\varepsilon}_{1, knot} (\mathbf{k}), \boldsymbol{\varepsilon}_{2, knot} (\mathbf{k}), \mathbf{k}\}$ form an orthogonal trihedron. As in the case of the $\epsilon^{\mu}_{\pm} (k)$ vectors, the new basis vectors $\varepsilon^{\mu}_{1,2} (k)$ contain the information about the topological properties of the electromagnetic field, but reduced due to the lesser number of parameters.

The electromagnetic tensor of the knot wave can be easily calculated and the following result is obtained
\begin{align}
F^{\mu \nu} (k) & = - 
\frac{e^{-\omega}}{\sqrt{2 \pi} \omega^{2}}
\left\{
	\left[
		k^{\mu}  \varepsilon^{\nu}_{2, knot} (k)
		-
		k^{\nu}  \varepsilon^{\mu}_{2, knot} (k)
	\right]
	\cos (k \cdot x)
\right.
\nonumber
\\
& +
\left[
		k^{\mu}  \varepsilon^{\nu}_{1, knot} (k)
		-
		k^{\nu}  \varepsilon^{\mu}_{1, knot} (k)
	\right]
	\sin (k \cdot x)
\left.
\right\}
\, .
\label{electromagnetic-tensor-knot}
\end{align}
From $F^{\mu \nu} (k)$ we can work out the Lorentz invariants of the real knot wave. We can show that both invariants vanish
as expected
\begin{align}
I_{1} (k) & = F_{\mu \nu} (k) F^{\mu \nu} (k) = 0
\, ,
\label{invariant-knot-wave-1}
\\
I_{2} (k) & = \tilde{F}_{\mu \nu} (k) F^{\mu \nu} (k) = 0
\, .
\label{invariant-knot-wave-2}
\end{align}
From this result we can conclude that the  real knot wave is a null field. 

In the above analysis, we have employed the helicity basis to express the gauge fixed potential $\mathsf{A}^{\mu} (x,k)$ as a linear combination of left and right circularly polarized waves. Due to the gauge and reference fixing, the electromagnetic knot field has only two degrees of freedom. By expressing these in the helicity basis, the topological properties become more transparent. For detailed discussions of these points we reefer to \cite{Ranada:1989wc, Ranada:1990,Ranada:1992hw,Ranada:1995,Arrayas:2011ci,Ranada:2017ore,Arrayas:2014,Arrayas:2018}. Similar considerations hold for the potential $\mathsf{C}^{\mu} (x, k )$.
The vector $\mathsf{A}^{\mu} (x, k )$ and $\mathsf{C}^{\mu} (x, k )$ are dual to each other. This duality can be used to emphasize either the topology of the magnetic lines by employing $\mathsf{A}^{\mu} (x, k )$
or the electric lines by using $\mathsf{C}^{\mu} (x, k )$. In either representation, the degrees of freedom are given by $ \mathbf{e}_{R} (\mathbf{k})$ and $\mathbf{e}_{L} (\mathbf{k})$, the left and right linearly independent helicity vectors, as described by the equations
(\refeq{epsilon-plus-App-A}) and (\refeq{epsilon-minus-App-A}). In what follows, we will focus on the systems with a classical background specified by the real electromagnetic potential $\mathsf{A}^{\mu} (x, k )$ from the equation  (\refeq{knot-mode-potential}) above.

\section{Classical particle in knot wave background}

In this section, we will analyse the dynamics of a classical particle of electric charge $q$ and mass $M$ that moves in the knot wave background presented above. To this end, we will use the Hamilton-Jacobi formalism in which the Hamilton action $S$ is a solution of the following equation
\begin{align}
& \left\{
\partial^{\mu} S +  
	\frac{q e^{-\omega}}{\sqrt{2 \pi} \omega^2}
			\left[
			\varepsilon^{\mu}_{1,knot} (k) \cos(k \cdot x)
			-
			\varepsilon^{\mu}_{2,knot} (k) \sin (k \cdot x)
			\right]
\right\}
\nonumber
\\
& \times
\left\{
\partial_{\mu} S +  
	\frac{q e^{-\omega}}{\sqrt{2 \pi} \omega^2}
			\left[
			\varepsilon_{1,knot, \, \mu} (k) \cos(k \cdot x)
			-
			\varepsilon_{2,knot, \, \mu} (k) \sin (k \cdot x)
			\right]
\right\}
= M^{2}
\, .
\label{Hamilton-Jacobi}
\end{align}
Since the knot wave has an explicit dependence on time, the system is non-autonomous. However, the field $\mathsf{A}^{\mu} (x, k )$ depends only on the Lorentz invariant phase $\phi = k \cdot x$. In this case, the Hamilton-Jacobi equation (\refeq{Hamilton-Jacobi}) can be solved exactly \cite{Landau:2004}. After some straightforward calculations, we obtain the following solution
\begin{align}
S_{p, \mathsf{A}} (\phi) & = - p_{in} \cdot x
- \frac{q e^{-\omega}}{\sqrt{2 \pi} \omega^2 p_{in} \cdot k}
	\left[
		p_{in} \cdot \varepsilon_{1,knot} (k) \sin( \phi )
		+ 
		p_{in} \cdot \varepsilon_{2,knot} (k) \cos (\phi )
	\right]
\nonumber
\\
& - \frac{q^{2} e^{- 2\omega}}{4 \pi \omega^2 p_{in} \cdot k}
\left[
	m^{2} \left( k^{2}_{1} + k^{2}_{2} \right)
	+
	s^{2} \left( k^{2}_{2} + k^{2}_{3} \right)
	+
	2 s m \omega k_{2}
\right] \phi + C_{in} (k)
\, , 
\label{Hamilton-Jacobi-solution}
\end{align}
where $p^{\mu}_{in}$ is the initial momentum of the particle $p^{\mu}_{in} = M u^{\mu}_{in}$, $u^{\mu}_{in}$ is the initial four-velocity and $C_{in} (k)$ is an integration constant that depends on the mode momentum and has the following explicit form
\begin{align}
C_{in} (k) & = - 
\frac{q e^{-\omega}}{\sqrt{2 \pi} \omega^2 p_{in} \cdot k}
	\left[
		p_{in} \cdot \varepsilon_{1,knot} (k) \sin( \phi_{in} )
		+
		p_{in} \cdot \varepsilon_{2,knot} (k) \cos (\phi_{in} )
	\right]
\nonumber
\\
& + \frac{q^{2} e^{- 2\omega}}{4 \pi \omega^2 p_{in} \cdot k}
\left[
	m^{2} \left( k^{2}_{1} + k^{2}_{2} \right)
	+
	s^{2} \left( k^{2}_{2} + k^{2}_{3} \right)
	+
	2 s m \omega k_{2}
\right] \phi_{in}
\, .
\label{Hamilton-Jacobi-integration-constant}
\end{align}
Without the loss of generality, we will set $C_{in} (k)$ to zero if convenes to do so. The canonical momentum of the particle is defined as
usual
\begin{equation}
P^{\mu} = - \partial^{\mu} S_{p, \mathsf{A}} (\phi) 
=  M u^{\mu} + 
\frac{q e^{-\omega}}{\sqrt{2 \pi} \omega^2}
		\left[
		\varepsilon^{\mu}_{1,knot} (k) \cos(\phi)
		- 
		\varepsilon^{\mu}_{2,knot} (k) \sin (\phi)
		\right]
\, .
\label{Hamilton-Jacobi-canonical-momentum}
\end{equation}
By solving the equation (\refeq{Hamilton-Jacobi-canonical-momentum}) for $p_{\mu}$, we obtain the following expression for the linear momentum of the particle
\begin{align}
p^{\mu} (\phi ) & = p^{\mu}_{in}
+
\frac{q^{2} e^{- 2\omega}}{4 \pi \omega^2 p_{in} \cdot k} k^{\mu}
\left[
	m^{2} \left( k^{2}_{1} + k^{2}_{2} \right)
	+
	s^{2} \left( k^{2}_{2} + k^{2}_{3} \right)
	+
	2 s m \omega k_{2}
\right]
\nonumber
\\
& - \frac{q e^{-\omega}}{\sqrt{2 \pi} \omega^2}
	\left\{
		\left[
			\varepsilon^{\mu}_{1,knot} (k) +
			k^{\mu} p_{in} \cdot \varepsilon_{1,knot} (k)
		\right]		
		\cos(\phi)
-
	\left[
			\varepsilon^{\mu}_{2,knot} (k) +
			k^{\mu} p_{in} \cdot \varepsilon_{2,knot} (k)
		\right] 
		\sin (\phi)
	\right\}
\, . 
\label{Hamilton-Jacobi-momentum}
\end{align}
The particle trajectory in space-time can be obtained by integrating $p^{\mu} ( \phi )$. The calculation is simple and it leads to the following result
\begin{align}
x^{\mu} (\phi )  & =  
x^{\mu}_{in} + \frac{p^{\mu}_{in}}{p_{in} \cdot k}
\left( \phi - \phi_{in} \right)
- \frac{q e^{-\omega}}{\sqrt{2 \pi} \omega^2 p_{in} \cdot k}
	\left[
		\varepsilon^{\mu}_{1,knot} (k) \sin( \phi )
		+
		\varepsilon^{\mu}_{2,knot} (k) \cos (\phi )
	\right]
\nonumber
\\
& + 
\frac{q e^{-\omega}}
{\sqrt{2 \pi} \omega^2 \left( p_{in} \cdot k \right)^2 } k^{\mu}
	\left[
		p_{in} \cdot \varepsilon_{1,knot} (k) \sin( \phi )
		+
		p_{in} \cdot \varepsilon_{2,knot} (k) \cos (\phi )
	\right]
\nonumber
\\
& -  
\frac{q^{2} e^{-2\omega}}{4 \pi \omega^2 \left( p_{in} \cdot k \right)^2}
k^{\mu}
\left[
	m^{2} \left( k^{2}_{1} + k^{2}_{2} \right)
	+
	s^{2} \left( k^{2}_{2} + k^{2}_{3} \right)
	+
	2 s m \omega k_{2}
\right] \phi
\, .
\label{Hamilton-Jacobi-trajectory}
\end{align} 
The equations (\refeq{Hamilton-Jacobi-momentum}) and (\refeq{Hamilton-Jacobi-trajectory}) have been obtained by ignoring the electromagnetic field of the particle, that is by considering a test particle. From the latter equation, we can see that the trajectory oscillates in all directions. In order to visualise the movement of the particle in the knot wave, it is convenient to split the equation (\refeq{Hamilton-Jacobi-trajectory}) into components. On the time-like direction, the trajectory has the following form
\begin{align}
x^{0} (\phi )  & =  
x^{0}_{in} + \frac{m u^{0}}{p_{in} \cdot k}
\left( \phi - \phi_{in} \right)
\nonumber
\\
& + 
\frac{q e^{-\omega}}
{\sqrt{2 \pi} \omega \left( p_{in} \cdot k \right)^2 }
	\left[
		p_{in} \cdot \varepsilon_{1,knot} (k) \sin( \phi )
		+
		p_{in} \cdot \varepsilon_{2,knot} (k) \cos (\phi )
	\right]
\nonumber
\\
& -  
\frac{q^{2} e^{-2\omega}}{4 \pi \omega \left( p_{in} \cdot k \right)^2}
\left[
	m^{2} \left( k^{2}_{1} + k^{2}_{2} \right)
	+
	s^{2} \left( k^{2}_{2} + k^{2}_{3} \right)
	+
	2 s m \omega k_{2}
\right] \phi
\, .
\label{Hamilton-Jacobi-trajectory-0}
\end{align} 
The spatial components of $x^{\mu} (\phi )$ can be conveniently decomposed 
into the sum of longitudinal and transversal vectors with respect to the knot wave momentum $\mathbf{k}$ as follows
\begin{align}
\mathbf{x} (\phi ) & = \mathbf{x}_{\parallel} (\phi ) 
+ \mathbf{x}_{\perp} (\phi )
\, ,
\label{Hamilton-Jacobi-trajectory-decomp-1}
\\
x^{j}_{\parallel} (\phi )  & = x_{\parallel} (\phi ) \frac{k^{j}}{\omega}
\, ,
\quad
x^{j}_{\perp} (\phi ) = x^{j}
-
k^{j} \frac{\mathbf{x}^{2} (\phi )}{\omega^2 }
\, ,
\quad j = 1, 2, 3 
\, .
\label{Hamilton-Jacobi-trajectory-decomp-2} 
\end{align}
After some simple algebra, we obtain for the longitudinal component the following expression
\begin{align}
x^{j}_{\parallel} (\phi ) & = 
\frac{\mathbf{k} \cdot \mathbf{x}_{in}}{\omega^2 }k^{j}
+ \frac{\mathbf{k} \cdot \mathbf{p}_{in}}{\omega^2 p_{in} \cdot k} k^{j}
\left( \phi - \phi_{in} \right)
\nonumber
\\
& + 
\frac{q e^{-\omega}}
{\sqrt{2 \pi} \omega^{2} \left( p_{in} \cdot k \right)^2 } k^{j}
	\left[
		p_{in} \cdot \varepsilon_{1,knot} (k) \sin( \phi )
		+
		p_{in} \cdot \varepsilon_{2,knot} (k) \cos (\phi )
	\right]
\nonumber
\\
& -  
\frac{q^{2} e^{-2\omega}}{4 \pi \omega^{2} \left( p_{in} \cdot k \right)^2} k^{j}
\left[
	m^{2} \left( k^{2}_{1} + k^{2}_{2} \right)
	+
	s^{2} \left( k^{2}_{2} + k^{2}_{3} \right)
	+
	2 s m \omega k_{2}
\right] \phi
\, .
\label{Hamilton-Jacobi-longitudinal-components}
\end{align}
The above equation shows that the longitudinal components of the trajectory undergo an oscillatory movement. More algebra leads to the following transversal vector
\begin{align}
x^{j}_{\perp} (\phi ) & = x^{j}_{in} + \frac{p^{j}_{in}}{p_{in} \cdot k}
\left( \phi - \phi_{in} \right)
- \frac{q e^{-\omega}}{\sqrt{2 \pi} \omega^2 p_{in} \cdot k}
	\left[
		\varepsilon^{j}_{1,knot} (k) \sin( \phi )
		+
		\varepsilon^{j}_{2,knot} (k) \cos (\phi )
	\right]
\nonumber
\\
& + 
\frac{q e^{-\omega}}
{\sqrt{2 \pi} \omega^2 \left( p_{in} \cdot k \right)^2 } k^{j}
	\left[
		p_{in} \cdot \varepsilon_{1,knot} (k) \sin( \phi )
		+
		p_{in} \cdot \varepsilon_{2,knot} (k) \cos (\phi )
	\right]
\nonumber
\\
& -  
\frac{q^{2} e^{-2\omega}}{4 \pi \omega^2 \left( p_{in} \cdot k \right)^2}
k^{j}
\left[
	m^{2} \left( k^{2}_{1} + k^{2}_{2} \right)
	+
	s^{2} \left( k^{2}_{2} + k^{2}_{3} \right)
	+
	2 s m \omega k_{2}
\right] \phi
\nonumber
\\
& - \frac{\mathbf{x}_{in}^{2}}{\omega^2} k^{j}
-
\frac{\mathbf{p}_{in}^{2}}{\omega^2 \left( p_{in} \cdot k\right)^2} 
\left( \phi - \phi_{in} \right)^{2}
k^{j}
\nonumber
\\
& - 
\frac{q^{2} e^{-2\omega}}{\pi \omega^4 \left( p_{in} \cdot k \right)^2} k^{j}
\left[
	m^{2} \left( k^{2}_{1} + k^{2}_{2} \right)
	+
	s^{2} \left( k^{2}_{2} + k^{2}_{3} \right)
	+
	2 s m \omega k_{2}
\right] \phi
\nonumber
\\
& -
\frac{q e^{- 2\omega}}
{2 \pi \omega^{4} \left( p_{in} \cdot k \right)^4 } k^{j}
	\left[
		p_{in} \cdot \varepsilon_{1,knot} (k) \sin( \phi )
		+
		p_{in} \cdot \varepsilon_{2,knot} (k) \cos (\phi )
	\right]^{2}
\nonumber
\\
& -
\frac{q^{4} e^{-4 \omega}}{16 \pi^{2} \omega^{4} \left( p_{in} \cdot k \right)^4} k^{j}
\left[
	m^{2} \left( k^{2}_{1} + k^{2}_{2} \right)
	+
	s^{2} \left( k^{2}_{2} + k^{2}_{3} \right)
	+
	2 s m \omega k_{2}
\right]^{2} \phi^{2}
\nonumber
\\
& -
\frac{2 \mathbf{x}_{in} \cdot \mathbf{p}_{in}}{\omega^2 p_{in} \cdot k} k^{j}
\left( \phi - \phi_{in} \right)
\nonumber
\\
& +
\frac{2 q e^{-\omega}}
{\sqrt{2 \pi} \omega^{4} \left( p_{in} \cdot k \right)^2 } k^{j}
	\left[
		\mathbf{x}_{in} \cdot \boldsymbol{\varepsilon}_{1,knot} (k) 
		\sin( \phi )
		+
		\mathbf{x}_{in} \cdot \boldsymbol{\varepsilon}_{2,knot} (k) 
		\cos (\phi )
	\right]
\nonumber
\\
& -
\frac{2 \mathbf{x}_{in}\cdot \mathbf{k} \, q e^{-\omega}}
{\sqrt{2 \pi} \omega^4 \left( p_{in} \cdot k \right)^2 } k^{j}
	\left[
		p_{in} \cdot \varepsilon_{1,knot} (k) \sin( \phi )
		+
		p_{in} \cdot \varepsilon_{2,knot} (k) \cos (\phi )
	\right]
\nonumber
\\
& + 
\frac{2 \mathbf{x}_{in}\cdot \mathbf{k} \, q^{2} e^{-2\omega}}{4 \pi \omega^4 \left( p_{in} \cdot k \right)^2} k^{j}
\left[
	m^{2} \left( k^{2}_{1} + k^{2}_{2} \right)
	+
	s^{2} \left( k^{2}_{2} + k^{2}_{3} \right)
	+
	2 s m \omega k_{2}
\right] \phi
\nonumber
\\
& +
\frac{2 q e^{-\omega}}
{\sqrt{2 \pi} \omega^4 \left( p_{in} \cdot k \right)^2 } k^{j}
	\left[
		\mathbf{p}_{in} \cdot \boldsymbol{\varepsilon}_{1,knot} (k) 
		\sin( \phi )
		+
		\mathbf{p}_{in} \cdot \boldsymbol{\varepsilon}_{2,knot} (k) 
		\cos (\phi )
	\right]	
	\left( \phi -\phi_{in} \right)
\nonumber
\\
& -
\frac{2 \mathbf{p}_{in}\cdot \mathbf{k} \, q e^{-\omega}}
{\sqrt{2 \pi} \omega^4 \left( p_{in} \cdot k \right)^3 } k^{j}
	\left[
		p_{in} \cdot \varepsilon_{1,knot} (k) 
		\sin( \phi )
		+
		p_{in} \cdot \varepsilon_{2,knot} (k) 
		\cos (\phi )
	\right]	
	\left( \phi -\phi_{in} \right)
\nonumber
\\
& +
\frac{2 \mathbf{p}_{in}\cdot \mathbf{k} \, q^2 e^{-2 \omega}}
{4 \pi \omega^4 \left( p_{in} \cdot k \right)^3 } k^{j}
\left[
	m^{2} \left( k^{2}_{1} + k^{2}_{2} \right)
	+
	s^{2} \left( k^{2}_{2} + k^{2}_{3} \right)
	+
	2 s m \omega k_{2}
\right] \phi \left( \phi -\phi_{in} \right)
\, .
\label{Hamilton-Jacobi-transversal-components}
\end{align}
Recall that the $x$-coordinates and $k$-coordinates are dimensionless. 
From the equations (\refeq{Hamilton-Jacobi-trajectory-0}), 
(\refeq{Hamilton-Jacobi-longitudinal-components}) and
(\refeq{Hamilton-Jacobi-transversal-components}), we can see that the longitudinal, transversal and temporal oscillations of the particle depend on the initial conditions $\mathbf{x}_{in}$ and $\mathbf{p}_{in}$.
These equations also show that the particle movement simplifies for certain initial values of $\mathbf{x}_{in}$ and $\mathbf{p}_{in}$. Let us set the initial data as well as $\phi_{in}$ to zero. Then the equations (\refeq{Hamilton-Jacobi-trajectory-0}), 
(\refeq{Hamilton-Jacobi-longitudinal-components}) and
(\refeq{Hamilton-Jacobi-transversal-components}) take the following simple form
\begin{align}
x^{0} (\phi )  & =  
\left[
\frac{u^{0}}{\omega} 
 -  
\frac{M^{2}_{\star} - M^2}{2 \omega M^2 } 
\right] \phi
\, ,
\label{Hamilton-Jacobi-trajectory-0-0}
\\
x^{j}_{\parallel} (\phi ) & = 
-  
k^{j} \frac{M^{2}_{\star} - M^2}{2 \omega^2 M^2 } \phi
\label{Hamilton-Jacobi-longitudinal-components-0}
\, ,
\\
x^{j}_{\perp} (\phi ) & = 
- \frac{q e^{-\omega}}{\sqrt{\pi} \omega^3 M}
	\left[
		\varepsilon^{j}_{1,knot} (k) \sin( \phi )
		+
		\varepsilon^{j}_{2,knot} (k) \cos (\phi )
	\right]
\nonumber
\\
& -  
k^{j} 
\left(
1
+ \frac{2}{\omega^2}
\right)
\left(
\frac{M^{2}_{\star} - M^2}{2 \omega^2 M^2 }
\right)
\phi
-
k^{j} \left( \frac{M^{2}_{\star} - M^2}{2 \omega^4 M^2 } \right)^2
\phi^{2}
\, ,
\label{Hamilton-Jacobi-transversal-components-0}
\end{align}   
where we have introduced the effective mass $M_{\star}$ by the following relation 
\begin{equation}
M^{2}_{\star}  = M^{2}
\left\{
1 +
\frac{q^{2} e^{-2\omega}}{2 \pi  M^2 \omega^2}
\left[
	m^{2} \left( k^{2}_{1} + k^{2}_{2} \right)
	+
	s^{2} \left( k^{2}_{2} + k^{2}_{3} \right)
	+
	2 s m \omega k_{2}
\right]
\right\}
\, .
\label{effective-mass-final}
\end{equation}
As the effective mass shows, in the trivial case $m = s = 0$ in which there is no knot wave the particle moves in the free space with its own mass $M$ as expected. As usual, $M_{\star}$ is also useful to calculate the time averaged observables in the periodic potential. 

The equations (\refeq{Hamilton-Jacobi-trajectory-0-0}) - (\refeq{Hamilton-Jacobi-transversal-components-0}) show that for vanishing initial conditions, the particle has a linear evolution along the time-like and parallel directions. On the time-like direction $x^{0}(\phi ) = 0$ for the frequency $\omega_{m,n} (\mathbf{k})$ given by the analytic continuations of the Lambert function
\begin{equation}
\omega_{m,n} (\mathbf{k})  = 
W_{N} 
\left(
\pm 
\frac{q}{M}\sqrt{\frac{1}{2 \pi u^{0}}
	\left[
	m^{2} \left( k^{2}_{1} + k^{2}_{2} \right)
	+
	s^{2} \left( k^{2}_{2} + k^{2}_{3} \right)
	+
	2 s m \omega k_{2}
	\right]}
\right)
\, ,
\quad N \in \mathbb{Z}
\, ,
\label{Hamilton-Jacobi-timelike-0-a}
\end{equation}
where
\begin{align}
& \frac{2 \pi M ^{2} u^{0}}{q^{2}} 
\left[
	m^{2} \left( k^{2}_{1} + k^{2}_{2} \right)
	+
	s^{2} \left( k^{2}_{2} + k^{2}_{3} \right)
	+
	2 s m \omega k_{2}
\right]^{-1}
\nonumber
\\
& \times
W_{N} 
\left(
\pm 
\frac{q}{M}\sqrt{\frac{1}{2 \pi u^{0}}
	\left[
	m^{2} \left( k^{2}_{1} + k^{2}_{2} \right)
	+
	s^{2} \left( k^{2}_{2} + k^{2}_{3} \right)
	+
	2 s m \omega k_{2}
	\right]}
\right) > 0
\, .
\label{Hamilton-Jacobi-timelike-0-b}
\end{align}
Since the Lambert function increases monotonically, for values of $\omega < \omega_{m,n} (\mathbf{k}) $ the coordinate $x^{0} (\phi )$ decreases with the increase of $\phi$. The opposite is true for $\omega >  \omega_{m,n} (\mathbf{k}) $. The frequency $\omega < \omega_{m,n} (\mathbf{k})$ takes a simpler value in the proper-time $s$  of the particle in terms of the Lambert function
\begin{equation}
\omega_{m,n} (\mathbf{k}; \, s )   = \frac{3}{2}
W
\left[
- \frac{2}{3}
\sqrt[3]
{
\frac{\pi \, p_{in} \cdot k}
{q^{2} M^{2}
\left[
	m^{2} \left( k^{2}_{1} + k^{2}_{2} \right)
	+
	s^{2} \left( k^{2}_{2} + k^{2}_{3} \right)
	+
	2 s m \omega k_{2}
\right]
}
}
\right]
\, .
\label{Hamilton-Jacobi-timelike-prop-time}
\end{equation} 
As we can see from the equation (\refeq{Hamilton-Jacobi-longitudinal-components-0}), the movement in the positive or negative projections $x^{j}_{\parallel} (\phi )$ along the axis of the longitudinal direction does not depend on the frequency of the wave, only on the sign of the components of $\mathbf{k}$ and on the invariant parameter $\phi$. The equation (\refeq{Hamilton-Jacobi-transversal-components-0}) shows that the zeroes of the coordinates in the transverse plane are determined by the following equation
\begin{equation}
U(k, \phi) \, \omega^{5} e^{-\omega} + \omega^{6} + 2 \omega^{4} + V (k,\phi) = 0
\, ,
\label{Hamilton-Jacobi-timelike-transverse-zeroes}
\end{equation} 
where the coefficients $U$ and $V$ are determined from the equation
(\refeq{Hamilton-Jacobi-transversal-components-0}). However, it is not clear if the solution of (\refeq{Hamilton-Jacobi-timelike-transverse-zeroes}) can be found analytically. 

The results obtained above can be used to calculate the observables of the system. One important observable is the energy scattered per unit frequency and solid angle. Its general formula in the far field approximation and in the relativistic form is \cite{Baier:1998vh}
\begin{equation}
\frac{d \mathcal{E}}{d \Omega^{\prime} d \omega^{\prime}} =
\frac{q^2 \omega^{\prime \, 2}}{4 \pi^{2}}
\left\vert 
\int^{+ \infty}_{- \infty} d \phi \, \frac{p^{\mu} (\phi)}{p_{in} \cdot k}
\exp\left[ i k^{\prime} \cdot x(\phi) \right]
\right\vert^{2}
\, ,
\label{Hamilton-Jacobi-covariant-density-energy}
\end{equation}
where $k^{\mu \, \prime} = ( \omega^{\prime} , \mathbf{k}^{\prime} )$ denotes the four-momentum of the emitted wave and $p^{\mu}_{in}$ is the four-momentum of the particle in the limit $\phi_{in} \to - \infty$. In order to apply the relation (\refeq{Hamilton-Jacobi-covariant-density-energy}) to our problem, we specify the direction of the emitted radiation and the initial conditions of the particle motion. Let us consider the emitted radiation along the knot wave, that is with $k^{\prime \, \mu} = a k^{\mu}$, where $a$ is a real positive with $a \neq 1$, by an electron with vanishing initial conditions $x^{\mu}_{in} = 0$, $\mathbf{p}_{in} = 0$ and $\phi_{in} = 0$. By using the equations (\refeq{Hamilton-Jacobi-momentum}) and (\refeq{Hamilton-Jacobi-trajectory}) into the formula (\refeq{Hamilton-Jacobi-covariant-density-energy}), we can solve analytically the integral from the right-hand side and we obtain after some lengthy calculations the following result
\begin{equation}
\frac{d \mathcal{E}_{\parallel}}{d \omega} =
q^2 a^{3}
\left[
\frac{\left( M_{\star} / M \right)^{2}}{a^2}
- \frac{\left( M_{\star} / M \right)^{2} - 1}{2 \pi \left(a^2 - 1 \right) \omega^2 }
\right]
\left[
1 + a^2 + \left( 1 - a^2 \right)
\cos \left( \left( a - 1 \right) \phi \right)
\cos \left( \left( a + 1 \right) \phi \right)
\right]
\, .
\label{Hamilton-Jacobi-covariant-density-energy-calc}
\end{equation}
Note that the limit $\phi_{in} = 0$ is purely artificial since $\phi > 0$ by definition. In the general case, the integral from the equation (\refeq{Hamilton-Jacobi-covariant-density-energy}) cannot be solved exactly due to the nonlinearities in the integrand and we expect that approximate and numerical methods be useful here.

\section{Quantization in a strong knot wave background}

In general, a particle of mass $M$ and electric charge $q$ interacts strongly with those electromagnetic modes for which $ \vert q \mathsf{A}_{knot} / M  \vert << 1$ \cite{Berestetsky:1982aq}. The amplitude of the knot is given by the first relation from (\refeq{knot-mode-potential}). That gives us an estimate for the range of frequencies in the case of knot waves
\begin{equation}
\omega \gg \omega_{cr} = - W 
\left[ 
-	\exp	\left(
			\frac{1}{2}
			\ln	\left(
			\frac{\vert q \vert}{\sqrt{2 \pi} M}
			\right)
			\right)
\right]
\, ,
\label{estimate-strong-frequencies}
\end{equation}
where $W [z]$ is the Lambert function. For the frequencies that satisfy this inequality, the corresponding electromagnetic knot field $\mathsf{F}_{\mu \nu} (x, k) = \partial_{\mu} \mathsf{A}_{\nu} (x, k) - \partial_{\nu} \mathsf{A}_{\mu} (x, k)$ can be treated as a strong classical background \cite{Berestetsky:1982aq}. In what follows, we are going to study the quantum dynamics of a spin half particle in the strong knot wave background by decomposing the total electromagnetic potential $A^{\mu}$ into the strong classical knot wave $\mathsf{A}^{\mu}$ and the quantum radiation field $\mathcal{A}^{\mu}$ at every point $x$ in space-time
\begin{equation}
A^{\mu} (x, k) = \mathsf{A}^{\mu} (x, k) + \mathcal{A}^{\mu} (x, k)
\, .
\label{decomposition-A}
\end{equation}
In this section, we are going to investigate the dynamics of the charged spin half particle in the field from the equation (\refeq{decomposition-A}) in the framework of the strong-field QED. We are going to quantize the system by applying the canonical quantization method. We will follow the general approach from \cite{Berestetsky:1982aq,Fradkin:1981sc}. (For a recent review of the strong-QED see \cite{Mackenroth:2014}.)

\subsection{Canonical quantization in the Furry picture}

In order to describe the effect of the intense knot wave field of frequency $\omega$ on a quantum spin half particle, we employ the Furry picture that was given for the first time in \cite{Furry:1951zz}. The Lagrangian density of the system is given by the following equation
\begin{equation}
\mathcal{L} =  
\bar{\psi} \left( i \slashed{\mathsf{D}} - m \right) \psi
- \frac{1}{4} \mathcal{F}^{2} - \frac{1}{2 \xi} \left( \partial \cdot \mathcal{A} \right)^{2}
- q \bar{\psi} \slashed{\mathcal{A}} \psi
\, ,
\label{lagrangian-total}
\end{equation}
where the covariant derivative in the knot wave field is 
\begin{equation}
\mathsf{D}^{\mu} = \partial^{\mu} + 
\frac{i q e^{-\omega}}{\sqrt{2 \pi} \omega^2}
\left[
			\varepsilon^{\mu}_{1,knot} (k) \cos(\phi)
			-
			\varepsilon^{\mu}_{2,knot} (k) \sin (\phi)
\right]
\, .
\label{covariant-derivative}
\end{equation}
Here, $\mathcal{F}_{\mu \nu} = \partial_{\mu} \mathcal{A}_{\nu} - \partial_{\nu} \mathcal{A}_{\mu}$ is the radiation field and $\xi$ is the gauge fixing parameter. As discussed above, we treat the knot wave as a non-dynamical background in which the Dirac field associated to the particle moves. The knot wave is the solution (\refeq{A-knot-mode-trig}) to the free Maxwell equations in vacuum for $\omega \gg \omega_{cr}$. The Dirac field $\psi$ and the radiation field $\mathcal{A}^{\mu}$ are solutions of the corresponding equations of motion in the presence of the $\mathsf{A}^{\mu}$ knot potential. 

After fixing the gauge, the system can be quantized by using the canonical quantization method \cite{Fradkin:1981sc,Furry:1951zz,Mackenroth:2014}. To this end, we impose the canonical commutation relations on the fields $\{ \psi, \bar{\psi}, \mathcal{A}^{\mu} \}$ and their canonically conjugate momenta $\{ \pi = i \psi^{\dagger}, \bar{\pi} = 0, \pi_{\mu} = - \frac{1}{4} \partial_{0} \mathcal{A}_{\mu} \}$. The corresponding Hamiltonian is the standard QED Hamiltonian in which the interaction with the knot wave is included in the \emph{free Hamiltonian}
\begin{equation}
H_{0} = H_{Dirac} + H_{Maxwell} + H_{int, \mathsf{A}}
\, .
\label{Hamiltonian-free}
\end{equation}
Here, we have denoted by $H_{Dirac}$ and $H_{Maxwell}$ the standard free Hamiltonians of the spin half and spin one fields. The interaction Hamiltonians with the knot wave and the radiation fields are
\begin{align}
H_{int, \mathsf{A}} & =  
\frac{q e^{-\omega}}{\sqrt{2 \pi} \omega^2}
\int d^3 \mathbf{x} \, \bar{\psi} (x)
\left[
			\slashed{\varepsilon}_{1,knot} (k) \cos( \phi)
			-
			\slashed{\varepsilon}_{2,knot} (k) \sin ( \phi )
\right]
\psi (x)
\, ,
\label{Hamiltonian-interaction-knot}
\\
H_{int, \mathcal{A}} & = q \int d^3 \mathbf{x} \, \bar{\psi} (x) \slashed{\mathcal{A}}^{\mu} \psi (x)
\, .
\label{Hamiltonian-interaction-photon}
\end{align}
As usual, we interpret the quantum fields $\{ \psi , \bar{\psi} ,\mathcal{A}^{\mu} \}$ as being formulated in the Schr\"{o}dinger picture. The transformations to the Furry picture are given by the following relations \cite{Fradkin:1981sc,Mackenroth:2014}
\begin{align}
O^{(F)} (t) & = U^{(F) \, \dagger}_{0} (t, t_{0}) \, O \, U^{(F)}_{0} (t, t_{0})
\, ,
\label{Furry-transformation-op}
\\
\vert \psi, \mathcal{A}; t \rangle^{(F)} & = U^{(F) \, \dagger}_{0} (t, t_{0})
\vert \psi, \mathcal{A}; t \rangle
\, ,
\label{Furry-transformation-state}
\\
U^{(F)}_{0} (t, t_{0}) & = T \exp\left[ - i \int^{t}_{t_{0}} d t^{\prime}H_{0} (t^{\prime}) \right]
\, ,
\label{Furry-unitary-operator}
\end{align}
where $O$ is an arbitrary operator in the Schr\"{o}dinger picture, the superscript $(F)$ indicates explicitly the corresponding objects in the Furry picture and the operator $U^{(F)}_{0} (t, t_{0})$ is unitary. It follows that the quantum dynamics is described by the following equations
\begin{align}
i \frac{d}{d t} O^{(F)}(t) & = \left[ O^{(F)}(t), H^{(F)}_{0}(t) \right]
\, ,
\label{quantum-equation-operators}
\\
i \frac{d}{d t} \vert \psi, \mathcal{A}; t \rangle^{(F)} & =
H^{(F)}_{int,\mathsf{A}}(t) \vert \psi, \mathcal{A}; t \rangle^{(F)}
\, ,
\label{quantum-equation-states}
\end{align}
where the state evolution operator is given by the mapping (\refeq{Furry-transformation-op}) applied to the interaction Hamiltonian with the radiation field
\begin{equation}
H^{(F)}_{int,\mathsf{A}}(t) = U^{(F) \, \dagger}_{0} (t, t_{0}) \, H_{int, \mathsf{A}}(t_{0}) \, U^{(F)}_{0} (t, t_{0})
\, .
\label{quantum-equation-state-evolution}
\end{equation}
The equations (\refeq{Furry-transformation-op})-(\refeq{quantum-equation-state-evolution}) show that the Furry picture is a particular case of the interaction picture in which the background was incorporated into the free Hamiltonian \cite{Berestetsky:1982aq}. By applying the relation (\refeq{quantum-equation-operators}) to the operators $\psi$ and $\mathcal{A}^{\mu}$, we obtain the following equations of motion of the field operators
\begin{align}
\left\{
i \slashed{\partial} - 
\frac{q e^{-\omega}}{\sqrt{2 \pi} \omega^2}
\left[
			\slashed{\varepsilon}_{1,knot} (k) \cos( \phi)
			-
			\slashed{\varepsilon}_{2,knot} (k) \sin ( \phi)
\right]
- M
\right\} \psi^{(F)} (x) & = 0
\, ,
\label{equation-motion-psi}
\\
\Box \, \mathcal{A}^{(F) \, \mu}  (x) & = 0
\, .
\label{equation-motion-A}
\end{align}
Next, we apply the Fourier transform to the above linear equations to find the mode decomposition of the field operators in the knot wave background. The fields have the following mode expansions
\begin{align}
\psi^{(F)}_{\mathsf{A}} (x) & =
\sum_{\alpha = 1, 2} \int d \mu(\mathbf{p})
\left[
c_{\mathbf{p},\alpha} \psi^{(F)}_{\mathsf{A},p,\alpha} (x)
+
d^{\dagger}_{\mathbf{p},\alpha} \psi^{(F)}_{\mathsf{A}, - p,\alpha} (x)
\right]
\, ,
\label{Fourier-decompostion-psi}
\\
\bar{\psi}^{(F)}_{\mathsf{A}} (x) & =
\sum_{\alpha = 1, 2} \int d \mu(\mathbf{p})
\left[
c^{\dagger}_{\mathbf{p},\alpha} \bar{\psi}^{(F)}_{\mathsf{A},p,\alpha} (x)
+
d_{\mathbf{p},\alpha} \bar{\psi}^{(F)}_{\mathsf{A}, - p,\alpha} (x)
\right]
\, ,
\label{Fourier-decomposition-bar-psi}
\\
\mathcal{A}^{(F) \, \mu} (x) & =
\sum_{\lambda} \int d \mu(\mathbf{k})
\left[
a_{\mathbf{k},\lambda} \varepsilon^{\mu}_{\lambda} 
e^{-ik \cdot x}
+
a^{\dagger}_{\mathbf{k},\lambda} \varepsilon^{\star \mu}_{\lambda} 
e^{ik \cdot x}
\right]
\, ,
\label{Fourier-decomposition-radiation-A}
\end{align}
where $\alpha$ are spin indices, $\varepsilon^{\mu}_{\lambda}$ and $\lambda$ are polarization vectors and indices, respectively, and $d \mu(\mathbf{p})$ and $d \mu(\mathbf{k})$ are the covariant measures in the corresponding phase spaces given by the equation (\refeq{covariant-measures-App-A}). In the Fourier decomposition of the fermionic field, the mode functions $\psi^{(F)}_{\mathsf{A},p,\alpha}$ are solutions of the Dirac equation (\refeq{equation-motion-psi}) in the presence of the strong knot wave $\mathsf{A}^{\mu}$. 

The states of the system, labelled by the pair $(\psi, \mathcal{A})$ or their mode operators, belong to the Fock space constructed from the vacuum state $\vert 0 \rangle$ in the presence of the knot wave background. The creation and annihilation operators that appear in the equations
(\refeq{Fourier-decompostion-psi})-(\refeq{Fourier-decomposition-radiation-A}) satisfy the canonical commutation relations
\begin{align}
\left[
c_{\mathbf{p},\alpha} , c^{\dagger}_{\mathbf{p^{\prime}},\alpha^{\prime}}
\right]_{+} 
& =
2 (2 \pi)^{3} \omega_{p} \, \delta^{3} (\mathbf{p} - \mathbf{p}^{\prime})
\delta_{\alpha , \alpha^{\prime}}
\, ,
\label{commutation-relations-c}
\\
\left[
d_{\mathbf{p},\alpha} , d^{\dagger}_{\mathbf{p^{\prime}},\alpha^{\prime}} 
\right]_{+} 
& =
2 (2 \pi)^{3} \omega_{p} \, \delta^{3} (\mathbf{p} - \mathbf{p}^{\prime})
\delta_{\alpha , \alpha^{\prime}}
\, ,
\label{commutation-relations-d}
\\
\left[
a_{\mathbf{k},\lambda} ,
a^{\dagger}_{\mathbf{k^{\prime}},\lambda^{\prime}} 
\right]_{-}
& = -
2 (2 \pi)^{3} \omega_{k} \, \delta^{3} (\mathbf{k} - \mathbf{k}^{\prime})
\eta_{\lambda , \lambda^{\prime}}
\, ,
\label{commutation-relations-a}
\end{align}
where $+/-$ stand for anticommutators and commutators, respectively, and the rest of the commutation relations are zero. As usual, we have obtained the relations (\refeq{commutation-relations-c})-(\refeq{commutation-relations-a}) from the equal time commutation relations of fields and their momenta. The vacuum state satisfies the equations
\begin{equation}
c_{\mathbf{p},\alpha} \vert 0 \rangle =
d_{\mathbf{p},\alpha} \vert 0 \rangle =
a_{\mathbf{k},\lambda} \vert 0 \rangle = 0
\, ,
\label{vacuum-state}
\end{equation}
for all values of $\mathbf{p}$, $k$, $\alpha$ and $\lambda$. A state that contains a number $f$ of particles of momentum $\mathbf{p}$, $g$ of antiparticles of momentum $\mathbf{p}^{\prime}$ and $h$ of photons of momentum 
$\mathbf{k}$ is denoted by
\begin{equation}
\vert
f_{\mathbf{p}}; g_{\mathbf{p}^{\prime}}; h_{\mathbf{k}}
\rangle
=
\left( c^{\dagger}_{\mathbf{p}}  \right)^{f}
\left( d^{\dagger}_{\mathbf{p}^{\prime}}  \right)^{g}
\left( a^{\dagger}_{\mathbf{k}}  \right)^{h}
\vert 0 \rangle 
\, .
\label{general-state}
\end{equation}
In the above relation, we have suppressed the polarization and spin indices and have defined the particles as being the quantum excitations with negative charge $-q$ as is the case of the electron. In general, it is known that the vacuum state in the presence of a strong background field suffers from some major problems \cite{Hartin:2018egj}. Namely, an intense field can excite a particle and leave the vacuum electrically charged and can close the gap between the particles and antiparticles. Since the knot wave has the same general properties as the plane waves, we can infer that the vacuum $\vert 0 \rangle$ in the presence of plane wave backgrounds is stable. 

\subsection{Volkov solutions and S-matrix}

The field $\mathcal{A}^{(F) \, \mu} (x)$ from the equation (\refeq{Fourier-decomposition-radiation-A}) already represents the solution of the free radiation field equation (\refeq{equation-motion-A}) in the Furry picture. However, the fields $\psi^{(F)}_{\mathsf{A}} (x)$ and $\bar{\psi}^{(F)}_{\mathsf{A}} (x)$ from the equations (\refeq{Fourier-decompostion-psi}) and (\refeq{Fourier-decomposition-bar-psi}) can be expressed in terms of linearly independent solutions of the Dirac equation (\refeq{equation-motion-psi}), known as \emph{Volkov solutions} \cite{Wolkow:1935zz}, only for a few types of background fields, among which there is the electromagnetic wave. 

The general form of the solutions of the equation (\refeq{equation-motion-psi}) can be obtained by using the properties from the relations (\refeq{Lorentz-gauge-k-A1-A2}) and the Appendix A. After somewhat lengthy but straightforward calculations that follow closely the general method from \cite{Fradkin:1981sc,Furry:1951zz,Mackenroth:2014},  we obtain the following Volkov states in the standard form
\begin{equation}
\psi^{(F)}_{\mathsf{A},p,\alpha} (x) = E_{p} (x) u_{p,\alpha}
\, ,
\label{Volkov-states-standard}
\end{equation}
where $ E_{p} (x)$ are the Ritus matrices \cite{Ritus:1985}
and $u_{p,\alpha}$ is the solution of the free on-shell Dirac equation
$\left( \slashed{p} - M \right) u_{p,\alpha} = 0$. It is useful to write the Ritus matrices as a product that can help with the algebraic manipulations and interpretations of the results, namely 
\begin{equation}
E_{p} (x) = \left\{ 1 + 
\frac{q e^{-\omega}}{2 \sqrt{2 \pi} \omega^2 p \cdot k}  \slashed{k}
\left[
			\slashed{\varepsilon}_{1,knot} (k) \cos(\phi)
			-
			\slashed{\varepsilon}_{2,knot} (k) \sin (\phi)
\right]
\right\}
\exp{i S_{p, \mathsf{A}}(x)}
\, ,
\label{Ritus-matrices-standard}
\end{equation}
where $S_{p, \mathsf{A}}(x)$ is the Hamilton function obtained in the previous section and given by the equation (\refeq{Hamilton-Jacobi-solution}) with the integration constant set to zero. The Volkov functions can be made more explicit in terms of the parameters of the wave. After some more algebra, we find that the Volkov solution in the knot wave background has the following analytic form
\begin{align}
& \psi^{(F)}_{\mathsf{A},p,\alpha} (x) = 
\left\{
1 + 
\frac{q e^{- \omega}}{\sqrt{8 \pi} \omega^{2} p \cdot k } \slashed{k}
	\left[
	m
	\left[
		\gamma_{3} \left( k^{2}_{1} + k^{2}_{2} \right)
		- \gamma_{1} k_{1} k_{3} - \gamma_{2} k_{2} k_{3}
	\right]
	+ s \omega 
		\left( 
			\gamma_{3} k_{2} - \gamma_{2} k_{3} 
		\right)
	\right] 
	\cos (\phi)	
\nonumber
\right.
\\
&-
\left.
\frac{q e^{- \omega}}{\sqrt{8 \pi} \omega^{2} p \cdot k } \slashed{k}
	\left[
		s
		\left[
		- \gamma_{1} \left( k^{2}_{2} + k^{2}_{3} \right)
		+ \gamma_{2} k_{1} k_{2} + \gamma_{3} k_{1} k_{3}
		\right]
		+ m \omega 
			\left( 
			\gamma_{2} k_{1} - \gamma_{1} k_{2} 
			\right)
		\right]	
		\sin (\phi)
	\right\}
\nonumber
\\
& \times
\exp
\left\{
- i p \cdot x 
- \frac{i q e^{-\omega}}{\sqrt{2 \pi} \omega^{2} p \cdot x}
	\left[
		m
		\left[
		p_{3} \left( k^{2}_{1} + k^{2}_{2} \right)
		- p_{1} k_{1} k_{3} - p_{2} k_{2} k_{3}
		\right]
		+ s \omega 
		\left( 
		p_{3} k_{2} - p_{2} k_{3} 
		\right) 
	\right] 
	\sin (\phi)
\right.
\nonumber
\\	
& +
\left.
\frac{i q e^{-\omega}}{\sqrt{2 \pi} \omega^{2} p \cdot x}
	\left[
		s
		\left[
		- p_{1} 
			\left( 
			k^{2}_{2} + k^{2}_{3} 
			\right)
		+ p_{2} k_{1} k_{2} + p_{3} k_{1} k_{3}
		\right]
		+ m \omega 
		\left( 
			p_{2} k_{1} - p_{1} k_{2} 
		\right)
	\right]
	\cos (\phi)
\right\}
\nonumber
\\
&
\times
\exp
\left\{
- i 
\left(
\frac{M ^{2}_{\star} - M^{2}}{2 \omega^{2} p \cdot k}
\right)
\phi
\right\}
u_{p,\alpha}
\, .
\label{Volkov-solution-knot-2}
\end{align}  
Some comments are in order here. Firstly, concerning the calculations, we note that the equation (\refeq{Volkov-solution-knot-2}) contains a new Volkov solution obtained in the particular case of a wave function, namely the real monochromatic knot. The wave property of the knot allows us to derive the Volkov solutions by following the general method for the plane wave which we have applied it as showed in \cite{Fradkin:1981sc,Furry:1951zz,Mackenroth:2014}. The only important detail worth noting here is the fact that we were able to determine the Volkov function of the sum of plane waves because each sine and cosine component of the wave satisfies independently the Lorentz gauge condition given by the equation  (\refeq{Lorentz-gauge-k-A1-A2}) above. Secondly, the main distinction between the Volkov state (\refeq{Volkov-solution-knot-2}) and a polarized plane wave resides in the information about the topological electromagnetic field that is expressed by the combinations of components of the wave momentum $\mathbf{k}$ and the real knot numbers $(m,s)$ that parametrize the functions $\psi^{(F)}_{\mathsf{A},p,\alpha} (x)$. As we have seen in the Section 2, these numbers are the two parameters allowed by the procedure of taking the real part of the complex knot mode and keeping orthogonal components. For $m = s = 0$, the functions $\psi^{(F)}_{\mathsf{A},p,\alpha} (x)$ describe a free field as expected
\begin{equation}
\psi^{(F) \, 0}_{\mathsf{A},p,\alpha} (x) = e^{- i p \cdot x} u_{p,\alpha}
\, .
\label{Volkov-solution-knot-3}
\end{equation}
One important advantage of knowing the Volkov states is that the set
$\{ \psi^{(F)}_{\mathsf{A},p,\alpha} \}_{p,\alpha}$ 
can be used to calculate the matrix elements of the scattering operator  $S^{(F)}_{\mathsf{A}}$ defined as usual by the following  relation \cite{Berestetsky:1982aq}
\begin{equation}
S^{(F)}_{\mathsf{A}}  = T 
\exp 
\left[ 
-i q \int d^{4} x 
: \bar{\psi}^{(F)}_{\mathsf{A}} (x)
\slashed{\mathcal{A}}^{(F)} (x)
\psi^{(F)}_{\mathsf{A}} (x) : 
\right]
\, ,
\label{S-operator}
\end{equation}
where the mode decomposition of the radiation field 
$\slashed{\mathcal{A}}^{(F)}$ is given by the equation (\refeq{Fourier-decomposition-radiation-A}) above. By applying the standard QFT methods, one can give the Dyson series of the operator $S^{(F)}_{\mathsf{A}}$ \cite{Itzykson:1980rh}
\begin{equation}
S^{(F)}_{\mathsf{A}} = \sum_{r = 1}^{\infty} \frac{(-1)^{s} q^{s}}{s !}
\int
d^{4} x_{1} \cdots d^{4} x_{s}
: \bar{\psi}^{(F)}_{\mathsf{A}} (x_{1})
\slashed{\mathcal{A}}^{(F)} (x_{1})
\psi^{(F)}_{\mathsf{A}} (x_{1}) :
\cdots
: \bar{\psi}^{(F)}_{\mathsf{A}} (x_{s})_{\mathsf{A}}
\slashed{\mathcal{A}}^{(F)} (x_{s})
\psi^{(F)}_{\mathsf{A}} (x_{s}) :
\, .
\label{S-operator-series}
\end{equation}
As usual, the relation (\refeq{S-operator-series}) can be used to calculate the scattering amplitudes of the quantum processes between the charged particles and the photons described by the states $\vert \psi , \mathcal{A} \rangle$ in the knot background $\mathsf{A}$.

\section{Compton effect in knot wave background}

As an example, let us consider the lowest order process of the nonlinear one-photon Compton scattering. The matrix element from the equation (\refeq{S-operator-series}) has the following form
\begin{equation}
S^{(F)}_{\mathsf{A}, fi} = 
\langle 
\mathbf{p}^{\prime}, \alpha^{\prime} ;
\mathbf{k}^{\prime}, \lambda^{\prime}
\vert
S^{(F)}_{\mathsf{A}}
\vert
\mathbf{p}, \alpha
\rangle
=
-i q \int d^{4} x \, \bar{\psi}^{(F)}_{\mathsf{A} , p^{\prime}, \alpha^{\prime}} (x)
\,
\slashed{\varepsilon}^{\star \, \prime}_{\lambda^{\prime}} 
\, 
\psi^{(F)}_{\mathsf{A},  p , \alpha} (x)  
\,
e^{i k^{\prime} \cdot x}
\, ,
\label{Compton-effect-S-matrix}
\end{equation}
where $(\mathbf{p}, \alpha)$ and $(\mathbf{p}^{\prime}, \alpha^{\prime})$ are the linear momentum and the spin of the initial and final particles, respectively, and $(\mathbf{k}^{\prime}, \lambda^{\prime})$ are the linear momentum and polarization of the emitted photon that is found in the final state of the process. 

The computation of the $S$-matrix from the equation (\refeq{Compton-effect-S-matrix}) is quite standard and it was used to discuss the Compton effect in a plane wave and a finite pulse backgrounds, see e. g. \cite{Boca:2011,Seipt:2016}. Nevertheless, let us outline its main points for clarity. By using the mode decompositions of fields given by the equations (\refeq{Fourier-decompostion-psi})-(\refeq{Fourier-decomposition-radiation-A}), we can write the $S$-matrix element of the Compton effect as follows
\begin{equation}
S^{(F)}_{\mathsf{A}, fi} =  
-i q 
\langle 0 \vert a_{\mathbf{k}^{\prime},\lambda^{\prime}} 
c_{\mathbf{p}^{\prime},\alpha^{\prime}}
\int d^{4} x
: \bar{\psi}^{(F)}_{\mathsf{A}} (x)
\slashed{\mathcal{A}}^{(F)} (x)
\psi^{(F)}_{\mathsf{A}} (x) : 
c^{\dagger}_{\mathbf{p},\alpha}
\vert 0 \rangle
\, .
\label{S-matrix}
\end{equation}
After performing the due contractions, we obtain the following result
\begin{align}
S^{(F)}_{\mathsf{A}, fi} & =
- i q \int d^{4} x \,
\bar{u}_{p^{\prime},\alpha^{\prime}}
\left\{ 1 + 
\frac{q e^{-\omega}}{2 \sqrt{2 \pi} \omega^2 p \cdot k}  \slashed{k}
\left[
			\slashed{\varepsilon}_{1,knot} (k) \cos(\phi)
			-
			\slashed{\varepsilon}_{2,knot} (k) \sin (\phi)
\right]
\right\}
\,
\slashed{\varepsilon}^{\star \, \prime}_{\lambda^{\prime}} 
\,
\nonumber
\\
& \times
\left\{ 1 + 
\frac{q e^{-\omega}}{2 \sqrt{2 \pi} \omega^2 p \cdot k}  \slashed{k}
\left[
			\slashed{\varepsilon}_{1,knot} (k) \cos(\phi)
			-
			\slashed{\varepsilon}_{2,knot} (k) \sin (\phi)
\right]
\right\}
u_{p,\alpha}
e^{i \left(S_{p}(x)- S_{p^{\prime}}(x) \right)}
\,
e^{i k^{\prime} \cdot x}
\, ,
\label{Compton-effect-S-matrix-1}
\end{align}
where $\varepsilon^{\star \, \prime}_{\lambda^{\prime}}$ is the polarization vector of the emitted photon and $S_{p}(x)$ is the Hamilton-Jacobi action with the particle momentum emphasized. By using the properties of the vectors $\varepsilon^{\mu}_{1, knot} (k)$ and $\varepsilon^{\mu}_{2, knot} (k)$ discussed in Section 2 and in the Appendix A and by performing some algebraic calculations, we can show that the phase function from the equation (\refeq{Compton-effect-S-matrix-1}) has the following form
\begin{align}
S_{p}(x)- S_{p^{\prime}}(x) & = 
\Gamma_{1} (p, p^{\prime}, k) \sin (\phi) +
\Gamma_{2} (p, p^{\prime}, k) \cos (\phi) +
\Gamma_{3} (p, p^{\prime}, k) \phi  
\, ,
\label{S-matrix-Gamma-function}
\end{align}
where the coefficients $\Gamma_{1,2,3} (p, p^{\prime}, k)$ are given by the following relations
\begin{eqnarray}
\Gamma_{1} (p, p^{\prime}, k)  & =  &
\frac{q e^{-\omega}}{\sqrt{2 \pi} \omega}
	\left[
	\frac{p^{\prime} \cdot \varepsilon_{1, knot} (k)}
	{p^{\prime} \cdot k} 
	- \frac{p \cdot \varepsilon_{1, knot} (k)}
	{p \cdot k}
	\right] 
\, ,
\nonumber
\\
\Gamma_{2} (p, p^{\prime}, k)  & =  & 
\frac{q e^{-\omega}}{\sqrt{2 \pi} \omega}
	\left[
	\frac{p^{\prime} \cdot \varepsilon_{2, knot} (k)}
	{p^{\prime} \cdot k} 
	- \frac{p \cdot \varepsilon_{2, knot} (k)}
	{p \cdot k}
	\right] 
\, ,
\nonumber
\\
\Gamma_{3} (p, p^{\prime}, k)  & = & 
\frac{ M^{2} \left(  M^{2}_{\star} - M^{2}\right) \omega^{4}}{4}
	\left[
	\frac{1}{p^{\prime} \cdot k} 
	- \frac{1}{ p \cdot k}
	\right] 
\, .
\label{G-coefficients}
\end{eqnarray}
The integrals from the right hand side of the equation (\refeq{Compton-effect-S-matrix-1}) are of the following general form
\begin{equation}
\Omega = \int d^{4} x F(\phi ) e^{i (k^{\prime} + p^{\prime} - p) \cdot x}
\, .
\label{Integrals-general-form}
\end{equation}
In order to solve them, it is convenient to pass to the light-front coordinates 
$ x^{\pm} = x^{0} \pm x^{3}$, $\mathbf{x}^{\perp} = \{ x^{1}, x^{2} \}$. Since the knot wave vector satisfies $k^{2} = 0$, its components can be chosen such that $k^{+} = 0$, $k^{-} = 2 \omega$, $\mathbf{k}^{\perp} = \mathbf{0}$ and the covariant phase is $\phi = \omega x^{+}$
\cite{Neville:1971zk}. It follows that the four-dimensional covariant integration measure in space-time is 
\begin{equation}
\sqrt{- \det({\eta_{\mu \nu})}} \, d^{4} x =  
\frac{1}{2 \omega} d \phi \, d x^{-} d^{2} \mathbf{x}^{\perp}
\, .
\label{Compton-effect-covariant-integration-measure}
\end{equation}
By integrating in the variables $x^{-}$ and $\mathbf{x}^{\perp}$, we obtain the following result
\begin{equation}
\Omega = \frac{\left( 2 \pi \right)^{3}}{\omega}
\delta^{(3)} 
\left( 
\mathbf{p} - \mathbf{p}^{\prime} - \mathbf{k}^{\prime}
\right)
\int^{+ \infty}_{- \infty} d \phi
F(\phi)e^{\frac{i}{2 \omega} (k^{\prime \, -} + p^{\prime \, -} - p^{-}) \phi }
\, .
\label{Integrals-just-phi}
\end{equation}
We introduce $\chi$ which is the conjugate variable to the phase $\phi$ 
and it is defined as follows
\begin{equation}
\chi = \frac{k^{\prime} \cdot p}{k \cdot p^{\prime}}
\, .
\label{conjugate-variable-chi}
\end{equation}
Then we can write the $S$-matrix as usual in terms of the transition amplitude
\begin{equation}
S^{(F)}_{\mathsf{A}, fi} =
- i q \frac{\left( 2 \pi \right)^{3}}{\omega}
\delta 
\left( 
p^{+} - p^{\prime \, +} - k^{\prime \, +}
\right)
\delta^{(2)} 
\left( 
\mathbf{p}_{\perp} - \mathbf{p}^{\prime}_{\perp} - \mathbf{k}^{\prime}_{\perp}
\right)
\mathcal{M} (\chi )
\, .
\label{Compton-effect-S-matrix-transition-amplitude}
\end{equation}
The integrals in $\phi$ from the equation (\refeq{Integrals-just-phi}) have the general form
\begin{equation}
I = \int^{+ \infty}_{- \infty} d \phi
f(\phi )
\exp 
\left\{ 	i 
	\left[
	\Gamma_{1} \sin (\phi ) + \Gamma_{2} \cos (\phi )
	+ \left( \Gamma_{3} + \chi \right) \phi
	\right] 
\right\}
\, ,
\label{Integrals-I}
\end{equation}
where $f(\phi)$ denotes the following functions
\begin{equation}
f(\phi)  = 
\{
1
\, ,
\cos (\phi)
\, ,
\sin (\phi)
\, ,
\cos^{2} (\phi)
\, ,
\sin (2 \phi)
\, ,
\sin^{2} (\phi)
\}
\, .
\label{functions-f}
\end{equation}
The integrals defined by the equations (\refeq{Integrals-I}) and (\refeq{functions-f}) cannot be solved directly due to the nonlinearities of their integrands. However, the integrands are periodic functions on $\phi$. Therefore, we can find analytic forms for these integrals in terms of series of Bessel functions \cite{Boca:2011}. In what follows, we are going to employ the following generating function \cite{Abramowitz:1965}
\begin{equation}
\exp \left[ \frac{z}{2} \left( t + \frac{1}{t} \right) \right]
=
\sum^{+ \infty}_{n = - \infty} t^{n} I_{n} (z)
\, , \quad t \neq 0 \, , z \in \mathbb{C}
\, ,
\label{Jacobi-Anger-relation}
\end{equation}
where $I_{n} (z)$ are the modified Bessel functions and the index $n$ must not be confused with the member of the quadruplet $(m, n, l, s)$ that label the knot. Then after some algebra and rearrangements, we obtain the following series representation of the transition matrix
\begin{align}
\mathcal{M} (\chi ) & = \sum_{n = - \infty}^{+ \infty}
\delta \left( n + \Gamma_{3} + \chi \right)
\left\{
	\left[
	2 \pi 
	\, \bar{u}_{p^{\prime} , \alpha^{\prime}}
	\, \slashed{\varepsilon}^{\star \, \prime}_{\lambda^{\prime}}
	\, u_{p,\alpha}
	+
	\frac{q^{2} e^{-\omega^{2}}}{8 \pi \omega^{4} (p \cdot k)^{2}} 
		\left[
		\bar{u}_{p^{\prime} , \alpha^{\prime}}
		\,
		\slashed{\varepsilon}_{1} 
		\,
		\slashed{k}
		\,
		\slashed{\varepsilon}^{\star \, \prime}_{\lambda^{\prime}}
		\,
		\slashed{k}
		\,
		\slashed{\varepsilon}_{1}
		\,
		u_{p,\alpha}
		\right.
	\right.
\right.
\nonumber
\\
& +
	\left.
		\left.
		\bar{u}_{p^{\prime} , \alpha^{\prime}}
		\,
		\slashed{\varepsilon}_{2} 
		\,
		\slashed{k}
		\,
		\slashed{\varepsilon}^{\star \, \prime}_{\lambda^{\prime}}
		\,
		\slashed{k}
		\,
		\slashed{\varepsilon}_{2}
		\,
		u_{p,\alpha}
		\right]
	\right]
	e^{ 
	- i n \arctan \left( \Gamma_{1}/\Gamma_{2} 
	\right)} 
	I_{n} \left( i \sqrt{\Gamma^{2}_{1} + \Gamma^{2}_{2}} \right)
\nonumber
\\
& +
\frac{q \sqrt{\pi} e^{- \omega}}{\sqrt{2} \omega^{2} p \cdot k}
		\left[
			k \cdot \varepsilon^{\star \, \prime}_{\lambda^{\prime}}
			\,
			\bar{u}_{p^{\prime} , \alpha^{\prime}}
			\left(			
			\slashed{\varepsilon}_{1}
			- i
			\slashed{\varepsilon}_{2}
			\right)			
			u_{p,\alpha}
			-
			\left(
			\varepsilon_{1} 
			- i
			\varepsilon_{2} 
			\right)
			\cdot 
			\varepsilon^{\star \, \prime}_{\lambda^{\prime}}
			\, 			
			\bar{u}_{p^{\prime} , \alpha^{\prime}}
			\,
			\slashed{k}
			\,
			u_{p,\alpha}
		\right]
\nonumber
\\
& \times
	e^{
	- i  (n - 1) \arctan \left( \Gamma_{1}/\Gamma_{2}
	\right)}
	I_{n - 1} \left( i \sqrt{\Gamma^{2}_{1} + \Gamma^{2}_{2}} \right)
\nonumber
\\
& +
\frac{q \sqrt{\pi} e^{- \omega}}{\sqrt{2} \omega^{2} p \cdot k}
		\left[
			k \cdot \varepsilon^{\star \, \prime}_{\lambda^{\prime}}
			\,
			\bar{u}_{p^{\prime} , \alpha^{\prime}}
			\left(			
			\slashed{\varepsilon}_{1}
			+ i
			\slashed{\varepsilon}_{2}
			\right)			
			u_{p,\alpha}
			-
			\left(
			\varepsilon_{1} 
			+ i
			\varepsilon_{2} 
			\right)
			\cdot 
			\varepsilon^{\star \, \prime}_{\lambda^{\prime}}
			\, 			
			\bar{u}_{p^{\prime} , \alpha^{\prime}}
			\,
			\slashed{k}
			\,
			u_{p,\alpha}
		\right]
\nonumber
\\
& \times
	e^{
	- i  (n + 1) \arctan \left( \Gamma_{1}/\Gamma_{2}
	\right)}
	I_{n + 1} \left( i \sqrt{\Gamma^{2}_{1} + \Gamma^{2}_{2}} \right)
\nonumber
\\
& +
\frac{q^{2} e^{- \omega^{2}}}{16 \omega^{4} (p \cdot k)^{2}}
		\left[
		\bar{u}_{p^{\prime} , \alpha^{\prime}} 
		\,
		\slashed{\varepsilon}_{1}
		\,
		\slashed{k}
		\,
		\slashed{\varepsilon}^{\star \, \prime}_{\lambda^{\prime}}
		\,
		\slashed{k}
		\,
		\slashed{\varepsilon}_{1}
		\,
		u_{p,\alpha}
		+ i
		\bar{u}_{p^{\prime} , \alpha^{\prime}}
		\,
		\slashed{\varepsilon}_{2}
		\,
		\slashed{k}
		\,\slashed{\varepsilon}^{\star \, \prime}_{\lambda^{\prime}}
		\,
		\slashed{k}
		\,
		\slashed{\varepsilon}_{1}
		\,
		u_{p,\alpha}
		\right.
\nonumber
\\
& + 
		i
		\left.
		\bar{u}_{p^{\prime} , \alpha^{\prime}}
		\,
		\slashed{\varepsilon}_{1}
		\,
		\slashed{k}
		\,
		\slashed{\varepsilon}^{\star \, \prime}_{\lambda^{\prime}}
		\,
		\slashed{k}
		\,
		\slashed{\varepsilon}_{2}
		\,
		u_{p,\alpha}
		-
		\bar{u}_{p^{\prime} , \alpha^{\prime}} 
		\,
		\slashed{\varepsilon}_{2}
		\,
		\slashed{k}
		\,
		\slashed{\varepsilon}^{\star \, \prime}_{\lambda^{\prime}}
		\,
		\slashed{k}
		\,
		\slashed{\varepsilon}_{2}
		\,
		u_{p,\alpha}
		\right]
	e^{
	- i  (n - 2) \arctan \left( \Gamma_{1}/\Gamma_{2}
	\right)}
	I_{n - 2} \left( i \sqrt{\Gamma^{2}_{1} + \Gamma^{2}_{2}} \right)
\nonumber
\\
& +
\frac{q^{2} e^{- \omega^{2}}}{16 \omega^{4} (p \cdot k)^{2}}
		\left[
		\bar{u}_{p^{\prime} , \alpha^{\prime}} 
		\,
		\slashed{\varepsilon}_{1}
		\,
		\slashed{k}
		\,
		\slashed{\varepsilon}^{\star \, \prime}_{\lambda^{\prime}}
		\,
		\slashed{k}
		\,
		\slashed{\varepsilon}_{1}
		\,
		u_{p,\alpha}
		- i
		\bar{u}_{p^{\prime} , \alpha^{\prime}}
		\,
		\slashed{\varepsilon}_{2}
		\,
		\slashed{k}
		\,\slashed{\varepsilon}^{\star \, \prime}_{\lambda^{\prime}}
		\,
		\slashed{k}
		\,
		\slashed{\varepsilon}_{1}
		\,
		u_{p,\alpha}
		\right.
\nonumber
\\
& - 
		i
		\left.
		\bar{u}_{p^{\prime} , \alpha^{\prime}}
		\,
		\slashed{\varepsilon}_{1}
		\,
		\slashed{k}
		\,
		\slashed{\varepsilon}^{\star \, \prime}_{\lambda^{\prime}}
		\,
		\slashed{k}
		\,
		\slashed{\varepsilon}_{2}
		\,
		u_{p,\alpha}
		-
		\bar{u}_{p^{\prime} , \alpha^{\prime}} 
		\,
		\slashed{\varepsilon}_{2}
		\,
		\slashed{k}
		\,
		\slashed{\varepsilon}^{\star \, \prime}_{\lambda^{\prime}}
		\,
		\slashed{k}
		\,
		\slashed{\varepsilon}_{2}
		\,
		u_{p,\alpha}
		\right]
	e^{
	- i  (n + 2) \arctan \left( \Gamma_{1}/\Gamma_{2}
	\right)}
	I_{n + 2} \left( i \sqrt{\Gamma^{2}_{1} + \Gamma^{2}_{2}} \right)
\left.
\right\}
\, .
\label{M-matrix-calculations}
\end{align}
Here, we have dropped the reference to the knot from the indices of the helicity operators and their arguments for simplicity. 
The relation (\refeq{M-matrix-calculations}) has the form
\begin{equation}
\mathcal{M} (\chi )  = \sum_{n = - \infty}^{+ \infty}
\delta \left( n + \Gamma_{3} + \chi \right) \mathcal{M}_{n} (\chi )
\, ,
\label{M-matrix-calculations-1}
\end{equation} 
in an obvious notation. Note that the relation $\mathcal{M} (\chi )$ is a series of $\delta$-functions similar to the one obtained in \cite{Boca:2011} in the case of a plane wave background. This results is expected since the knot wave has similar properties to the plane wave with the exception of the vectors $\varepsilon^{\mu}_{1,2}$ that contain information about the topological properties of the electromagnetic knot. A direct consequence of the $\delta$-functions is that the difference between the allowed values of the previously continuous variable $\chi$ is now discrete. 

The calculation of the emission probability can be done by interpreting the four momentum conservation as being the finite space-time volume $VT$
\begin{equation}
\delta^{(4)}(p_{in} - p_{fin}) \delta^{(4)}(0) = \frac{VT}{(2 \pi)^{4}} \delta ( p_{in} - p_{fin} )
\, .
\label{approximation-VT}
\end{equation}
The relevant quantity in this case is the differential rate of the quantum transition per unit time $d \dot{W} \simeq \vert S^{(F)}_{\mathsf{A}, fi} \vert^{2} /T$. From the equation (\refeq{Compton-effect-S-matrix-transition-amplitude}) we can see that 
\begin{equation}
\vert S^{(F)}_{\mathsf{A}, fi} \vert^{2} =
\frac{\left( 2 \pi \right)^{6} q^{2}}{\omega^{2}}
\delta 
\left( 
0
\right)
\delta^{(3)} 
\left( 
\mathbf{p} - \mathbf{p}^{\prime} - \mathbf{k}^{\prime}
\right)
\vert \mathcal{M}\vert^{2}
\, ,
\label{Compton-effect-S-matrix-transition-square}
\end{equation}
where we are using a simplified notation for the three-dimensional momentum delta-functions in the light-front coordinates. This is the same convention as the one used in \cite{Seipt:2010ya,Seipt:2016} to calculate the transition amplitudes in the presence of plane waves and laser pulses. The total and the partial cross sections for different transitions are calculated by averaging over the initial and final spin states $\alpha$ and $\alpha^{\prime}$, namely
\begin{equation}
\vert \mathcal{M} \vert^{2} = \frac{1}{2} \sum_{\alpha} \sum_{\alpha^{\prime}} \vert \mathcal{M} (\chi ) \vert^{2} 
\, .
\label{spin-average}
\end{equation}
The simplest process from $\mathcal{M} (\chi )$ is $\bar{u} \, \slashed{\varepsilon}^{\star \, \prime} u$ which does not involve the knot field photons that determine only the exponential and the Bessel function factor. The contribution of the knot field photons start to show up in the processes $\bar{u}\left(\slashed{\varepsilon}_{1} \mp i \slashed{\varepsilon}_{2} \right) u$ and $\bar{u}\, \slashed{k} \, u$ in which the Dirac particle interacts only with the background photons. The more complicated interactions like 
$\bar{u}\,\slashed{\varepsilon}_{A}\,\slashed{k}\,\slashed{\varepsilon}^{\star \, \prime}\,\slashed{k}\,\slashed{\varepsilon}_{B}\,u$, where $A, B = 1, 2$ involve all the degrees of freedom of the Dirac particle, the knot field photons and the emitted photon. 

In order to get a feeling of the interaction with the knot mode, let us work out the probabilities of some partial transition amplitudes in the general case. By using the properties of the Dirac matrices, we calculate the following probabilities for processes that do not involve the knot photons
\begin{align} 
\vert \mathcal{M}_{n, \varepsilon^{\prime}}  (\chi ) \vert^{2}
& =
\vert \bar{u} \, \slashed{\varepsilon}^{\star \, \prime} u \vert^{2}
= 8 \pi^{2} 
\left[
	\left(
	\varepsilon^{\star \, \prime}_{\lambda^{\prime}} \cdot p^{\prime}
	\right)
	\left( 
	\varepsilon^{\prime}_{\lambda^{\prime}} \cdot p
	\right)
+
	\left(
	\varepsilon^{\star \, \prime}_{\lambda^{\prime}} \cdot p
	\right)
	\left( 
	\varepsilon^{\prime}_{\lambda^{\prime}} \cdot p^{\prime}
	\right)
\right.
\nonumber
\\
& -
\left. 
	\left(
	p^{\prime} \cdot p - M^{2} 
	\right)
	\, 
	\varepsilon^{\star \, \prime}_{\lambda^{\prime}}
	\cdot
	\varepsilon^{\prime}_{\lambda^{\prime}}
\right]
\vert
I_{n} \left( i \sqrt{\Gamma^{2}_{1} + \Gamma^{2}_{2}} \right)
\vert^{2}
\, ,
\label{probability-bar-u-epsilon-u}
\\
\vert \mathcal{M}_{n, k}  (\chi ) \vert^{2} 
& =
\vert \bar{u} \, \slashed{k} u \vert^{2}
= \frac{\pi^{2} q^{2} e^{-2 \omega}}{2 \omega^{4}} 
\frac{k \cdot p^{\prime}}{k \cdot p}
\vert
I_{n} \left( i \sqrt{\Gamma^{2}_{1} + \Gamma^{2}_{2}} \right)
\vert^{2}
\, ,
\label{probability-bar-u-k-u}
\end{align}
Here, $\varepsilon^{\prime}_{\lambda^{\prime}}$ is the complex conjugate of $\varepsilon^{\star \, \prime}_{\lambda^{\prime}} $ and the overall coefficients from (\refeq{M-matrix-calculations}) have been restored.
The first processes in which the knot photon interacts with the Dirac particle have the following probabilities
\begin{align}
\vert \mathcal{M}_{n, \pm}  (\chi ) \vert^{2}
& =
\vert 
\bar{u}
\left(
\slashed{\varepsilon}_{1} \mp i \slashed{\varepsilon}_{2} 
\right) u 
\vert^{2}
\nonumber
\\
& = 
\frac{\pi^{2} q^{2} e^{-2 \omega}}{2 \omega^{4} (p \cdot k)^{2}} 
\left\{
	\left[
	 - m k_{1} k_{3} p_{1} 
	 -	\left( 
	 		m k_{2} k_{3} - s \omega k_{3}
	 	\right) p_{2}
	 + \left(
	 	m 
	 	\left( 
	 	k^{2}_{1} + k^{2}_{2}
 		\right)
 		- s \omega k_{2} 
 		\right) p_{3}
	\right]
\right.
\nonumber
\\
& \times
\left.
	\left[
	 - m k_{1} k_{3} p^{\prime}_{1} 
	 -	\left( 
	 		m k_{2} k_{3} - s \omega k_{3}
	 	\right) p^{\prime}_{2}
	 + \left(
	 	m 
	 	\left( 
	 	k^{2}_{1} + k^{2}_{2}
 		\right)
 		- s \omega k_{2} 
 		\right) p^{\prime}_{3}
	\right]
\right.
\nonumber
\\
& +
\left.
	\left[
	 \left( m \omega k_{2} 
	 + s	\left( k^{2}_{2} + k^{2}_{3} 
	 		\right)
	 \right) p_{1}
	 -	\left( 
	 		m \omega k_{1} + s k_{1} k_{2}
	 	\right) p_{2} 
	 - s \omega k_{2} k_{3} p_{3}
	\right]
\right.
\nonumber
\\
& \times
\left.
	\left[
	 \left( m \omega k_{2} 
	 + s	\left( k^{2}_{2} + k^{2}_{3} 
	 		\right)
	 \right) p^{\prime}_{1}
	 -	\left( 
	 		m \omega k_{1} + s k_{1} k_{2}
	 	\right) p^{\prime}_{2} 
	 - s \omega k_{2} k_{3} p^{\prime}_{3}
	\right]
\right.
\nonumber
\\
&+
\left.
\frac{2 \pi M^{2} \left(M^{2}_{\star} - M^{2}\right) \omega^{4}}{q^{2}e^{-2 \omega}}
	\left(
	p^{\prime} \cdot p  -  M^{2}
	\right)
\right\}
\vert
I_{n \mp 1} \left( i \sqrt{\Gamma^{2}_{1} + \Gamma^{2}_{2}} \right)
\vert^{2}
\, .
\label{probability-bar-u-epsilon-pm-u}
\end{align}
In the above expression, we have used the explicit form of the vectors 
$\varepsilon^{\mu}_{1, knot}$ and $\varepsilon^{\mu}_{2, knot}$ given by the relations (\refeq{knot-amplitude-epsilons}) as discussed in Section 2. 
In the equations (\refeq{probability-bar-u-epsilon-u}), (\refeq{probability-bar-u-k-u}) and (\refeq{probability-bar-u-epsilon-pm-u}), the background mode is present in the argument of the Bessel function and of the exponential. Their explicit form is given in the Appendix B. 

The complexity of the partial transition amplitudes increases when more processes are considered. For example, the real part of the partial transition amplitudes 
$\mathcal{M}_{n, \varepsilon^{\prime}}  (\chi ) 
\mathcal{M}^{\star}_{n, \mp}  (\chi )$
has the  following expression
\begin{equation}
\Re \left[ \mathcal{M}_{n, \varepsilon^{\prime}}  (\chi ) 
\mathcal{M}^{\star}_{n, \mp}  (\chi ) \right] = 
\Re \left[ \mathcal{M}_{n, \varepsilon^{\prime}}  (\chi ) 
\mathcal{M}^{\star}_{n, \mp}  (\chi ) \right]_{1}
+
\Re \left[ \mathcal{M}_{n, \varepsilon^{\prime}}  (\chi ) 
\mathcal{M}^{\star}_{n, \mp}  (\chi ) \right]_{2}
\, .
\label{real-part-M-M-general}
\end{equation}
The two terms are quite large expressions in terms of $(m,s)$ which are given in the Appendix B. The other processes from the relation (\refeq{M-matrix-calculations}) are also massively large due to the trace of the Dirac-matrices that generate 960 terms for each product and for arbitrary values of momenta.

\subsection{Partial transition amplitudes in particular cases}

Let us discuss in more detail the partial transition amplitudes
$ \vert \mathcal{M}_{n, \varepsilon^{\prime}}  (\chi ) \vert^{2}$,
$\vert \mathcal{M}_{n, k}  (\chi ) \vert^{2}$ and $\vert \mathcal{M}_{n, \pm}  (\chi ) \vert^{2}$ for some particular knot backgrounds and fixed particle and photon states. As a general idea, in order to pick up a concrete knot background, one has to fix its topological properties by choosing values for the knot numbers $(m,s)$ and its physical properties as energy and wave vector by defining the values of 
$k^{\mu} = (\omega, \mathbf{k})$. This amounts to fixing six real parameters. On the other hand, the quantum states are defined in terms of the ingoing and outgoing particles four momenta $p^{\mu}$ and $p^{\prime \mu}$, respectively, which gives us other eight parameters. Finally, by fixing the values of 
$\varepsilon^{\star \prime \mu}_{\lambda^{\prime}}$ and by determining $k^{\prime \mu}$ from the conservation laws, one defines the properties of the emitted photon. Recall that the reference frame was already fixed at the beginning together when the gauge was fixed as well.
By taking into account all these parameters, the first non-trivial knot waves are defined by the pairs $(m = 0, s= 1)$ and $(m=1, s=0 )$. It is instructive to discuss these two knot backgrounds as they provide an example of the statement that the partial transition amplitudes in different backgrounds and for different states are related by an equivalence relation, in the sense that the functions that describe the corresponding probabilities are the same. 

For simplicity, we will calculate explicitly only the partial processes with $n = 0$ from the relation (\refeq{M-matrix-calculations}). To this end, we introduce the notation $\zeta = \sqrt{\Gamma^{2}_{1} + \Gamma^{2}_{2}} > 0$ for the variable of the Bessel functions.

\subsubsection{Case $m = 0, s = 1$, $n = 0$}

In this case, we fix the rest of the parameters of the states for which we calculate the partial amplitudes as follows
\begin{align}
k_{\mu}	&	=	\left( k_{0} , k_{1}, k_{2}, k_{3} \right) 
			=	\left( \omega , 0, 0, k_{3} = k \right)
\, ,
\nonumber
\\
\varepsilon^{\mu}_{1} & = \left( 0, \boldsymbol{\varepsilon}_{1} \right)
			=	\left( 0, 0, \omega k, - \omega k \right)
\nonumber
\\
\varepsilon^{\mu}_{2} & = \left( 0, \boldsymbol{\varepsilon}_{2} \right)
			=	\left( 0, -k^{2}, 0, 0 \right)
\nonumber
\\  
p_{\mu}	&	=	\left( p_{0} , p_{1}, p_{2}, p_{3} \right) 
			=	\left( \omega_{p} , 0, p_{2} = p, 0 \right)
\, ,
\nonumber
\\
p^{\prime}_{\mu}	&	=	\left( p^{\prime}_{0} , p^{\prime}_{1}, 
							p^{\prime}_{2}, p^{\prime}_{3}
						\right) 
					=	\left( \omega_{p^{\prime}}, 0, p^{\prime}_{2} =
						p^{\prime} , 0 
						\right)
\, ,
\nonumber
\\
\varepsilon^{\star \prime}_{\lambda^{\prime}, 1, \mu} & = 
	\left( 
	\varepsilon^{\star \prime \mu}_{\lambda^{\prime}, 1, 0}, 
	\boldsymbol{\varepsilon}^{\star \prime}_{\lambda^{\prime} , 1} 
	\right)
	= \left( 0, 0, 1, 0\right)
\nonumber
\\
\varepsilon^{\star \prime}_{\lambda^{\prime}, 2, \mu} & = 
	\left( 
	\varepsilon^{\star \prime \mu}_{\lambda^{\prime}, 2, 0}, 
	\boldsymbol{\varepsilon}^{\star \prime}_{\lambda^{\prime} , 2} 
	\right)
	= \left( 0, 0, 0, 0 \right)
\, ,				
\label{fixed-states}
\end{align}
with
\begin{equation}
\varepsilon^{\star \, \prime}_{\lambda^{\prime}, \mu} = \varepsilon^{\star \, \prime}_{\lambda^{\prime}, 1 , \mu} + i \varepsilon^{\star \, \prime}_{\lambda^{\prime}, 2 , \mu}
\, .
\label{varepsilon-star-decomposed}
\end{equation}
Note that for the emitted photon we have 
\begin{align}
- \omega^{\prime} + k^{\prime}_{3} & = \omega_{p^{\prime}} - \omega_{p}
+
p_{3} - p^{\prime}_{3}
\, ,
\label{emitted-photon-delta-fct-1}
\\
\mathbf{k}^{\prime}_{\perp} & = \mathbf{p}_{\perp} - \mathbf{p}^{\prime}_{\perp}
\, .
\label{emitted-photon-delta-fct-2}
\end{align}
The above relations are conservation laws in the presence of the knot background and they are a consequence of the light-front delta functions. By using the equation (\refeq{fixed-states}) - (\refeq{emitted-photon-delta-fct-2}) and the energy-momentum relation for the incoming and outgoing particles $\omega^{2} = p^{2} + M^{2}$ together with the dispersion relations $\omega = \vert \mathbf{k} \vert$ for the photons, we can express the frequency of the emitted photon in terms of the energies of the incoming and outgoing particles as follows
\begin{equation} 
\omega^{\prime} = 
\frac{\omega^{2}_{p} + \omega^{2}_{p^{\prime}} - \omega_{p} \omega_{p^{\prime}} - M^{2} - \sqrt{\left(\omega^{2}_{p} - M^{2} \right)
\left(\omega^{2}_{p^{\prime}} - M^{2} \right)}}{\omega_{p} - \omega_{p^{\prime}}}
\, .
\label{emitted-photon-energy-Case-1}
\end{equation}
The partial amplitudes of the states from the equations (\refeq{fixed-states}) can be easily calculated from the relations (\refeq{probability-bar-u-epsilon-u}), (\refeq{probability-bar-u-k-u}) and (\refeq{probability-bar-u-epsilon-pm-u}). After some simple algebra, we get the following expressions for the corresponding probabilities
\begin{align} 
\vert \mathcal{M}_{0, \varepsilon^{\prime}}  (\chi ) \vert^{2}
& = 8 \pi^{2} 
\left(
	\omega_{p}\omega_{p^{\prime}} + p^{\prime} p - M^{2} 
\right)
\vert
J_{0} \left( \sqrt{\Gamma^{2}_{1} + \Gamma^{2}_{2}} \right)
\vert^{2}
\, ,
\label{probability-bar-u-epsilon-u-case-1}
\\
\vert \mathcal{M}_{0, k}  (\chi ) \vert^{2} 
& =
\frac{\pi^{2} q^{2} \omega_{p^{\prime}}}{2 \omega_{p}} 
\frac{ e^{-2 \omega}}{\omega^{4}}
\vert
J_{0} \left( \sqrt{\Gamma^{2}_{1} + \Gamma^{2}_{2}} \right)
\vert^{2}
\, ,
\label{probability-bar-u-k-u-case-1}
\\
\vert \mathcal{M}_{0, \pm}  (\chi ) \vert^{2}
& = 
\frac{\pi^{2} q^{2}}{2 \omega_{p}^{2}} 
\left(
	\omega_{p^{\prime}} \omega_{p} -   M^{2}
\right)
\frac{e^{-2 \omega}}{\omega^{2}}
\vert
J_{\mp 1} \left( \sqrt{\Gamma^{2}_{1} + \Gamma^{2}_{2}} \right)
\vert^{2}
\, .
\label{probability-bar-u-epsilon-pm-u-case-1}
\end{align}
Due to our fixing of parameters, we have obtained in the above relations the probability amplitudes as functions of the frequency of the monochromatic wave $\omega$. The only parameters left are the charge and rest mass of the incoming and outgoing particles as well as their energies and momenta. Note that the Bessel functions are also $\omega$ dependent because their argument takes the following form for the chosen states
\begin{align}
\zeta^{2} = \Gamma^{2}_{1} + \Gamma^{2}_{2}
& =
\frac{q^{2} e^{- 2 \omega}}{2 \pi}
\left(
\frac{p^{\prime}}{\omega_{p^{\prime}}}
-
\frac{p}{\omega_{p}}
\right)^{2}
\, .
\label{Gamma-1-Gamma-2-fin-2-particular-1}
\end{align}
Also, the effective mass defined by the equation (\refeq{effective-mass-final}) is $\omega$ dependent and it is given by the following relation
\begin{equation}
M^{2}_{\star}  = M^{2}
\left(
1 +
\frac{q^{2} e^{-2\omega}}{2 \pi  M^2}
\right)
\, .
\label{effective-mass-final-case-1}
\end{equation}
Even in this relatively simple case, the dependence on the probability amplitudes on the knot frequency is quite complicate due to the Bessel functions. Therefore, it is instructive to consider the following two limits.

\paragraph{I) $\zeta \to 0$} 
For large values of the frequency of the knot wave background the argument of the Bessel functions go to zero. By using the relations collected in the Appendix C, we obtain the following asymptotic expressions of the probabilities
\begin{align} 
\vert \mathcal{M}_{0, \varepsilon^{\prime}}  (\chi ) \vert^{2}
& \simeq 8 \pi^{2} 
\left(
	\omega_{p}\omega_{p^{\prime}} + p^{\prime} p - M^{2} 
\right)
\, ,
\label{probability-bar-u-epsilon-u-case-1-num-0}
\\
\vert \mathcal{M}_{0, k}  (\chi ) \vert^{2} 
& \simeq
\frac{\pi^{2} q^{2} \omega_{p^{\prime}}}{2 \omega_{p} } 
\frac{e^{-2 \omega}}{ \omega^{4}}
\, ,
\label{probability-bar-u-k-u-case-1-num-0}
\\
\vert \mathcal{M}_{0, \pm}  (\chi ) \vert^{2}
& \simeq 
\frac{ q^{4}}{32 \omega_{p}^{2}} 
\left(
	\omega_{p^{\prime}} \omega_{p} -   M^{2}
\right)
\left[
\left(
\frac{p^{\prime}}{\omega_{p^{\prime}}}
\right)^{2}
+
\left(
\frac{p}{\omega_{p}}
\right)^{2}
\right]^{2}
\frac{e^{- 6 \omega}}{\omega^{2}}
\, .
\label{probability-bar-u-epsilon-pm-u-case-1-num-0}
\end{align}
Note that the probabilities of these partial amplitudes decrease with the increasing of the frequency, with the exception of the contribution of the polarization which is constant.

\paragraph{II) $\zeta \to \infty$} 
The limit $\zeta = \sqrt{\Gamma^{2}_{1} + \Gamma^{2}_{2}} \to \infty$ is not achieved for any value of the frequency $\omega \in (0, \infty )$. However, for any fix $\omega$, the processes at low energies $\omega_{p}, \omega_{p^{\prime}} \to \infty$ generate large values of the argument of the Bessel functions. It follows that the probabilities are described by the following asymptotic expressions
\begin{align} 
\vert \mathcal{M}_{0, \varepsilon^{\prime}}  (\chi ) \vert^{2}
& \simeq 16 \pi 
\left(
	\omega_{p}\omega_{p^{\prime}} + p^{\prime} p - M^{2} 
\right)
\left\{
\frac{q^{2} e^{- 2 \omega}}{2 \pi }
\left[
\left(
\frac{p^{\prime}}{\omega_{p^{\prime}}}
\right)^{2}
+
\left(
\frac{p}{\omega_{p}}
\right)^{2}
\right]
\right\}^{-1}
\nonumber
\\
& \times
\cos^{2} \left[ 
\frac{q^{2} e^{- 2 \omega}}{2 \pi }
\left[
\left(
\frac{p^{\prime}}{\omega_{p^{\prime}}}
\right)^{2}
+
\left(
\frac{p}{\omega_{p}}
\right)^{2}
\right]
-
\frac{\pi }{4} \right]
\, ,
\label{probability-bar-u-epsilon-u-case-1-num-ifty}
\\
\vert \mathcal{M}_{0, k}  (\chi ) \vert^{2} 
& \simeq
\frac{\pi q^{2} \omega_{p^{\prime}}}{\omega_{p}} 
\frac{e^{-2 \omega}}{\omega^{4}}
\left\{
\frac{q^{2} e^{- 2 \omega}}{2 \pi }
\left[
\left(
\frac{p^{\prime}}{\omega_{p^{\prime}}}
\right)^{2}
+
\left(
\frac{p}{\omega_{p}}
\right)^{2}
\right]
\right\}^{- 1}
\nonumber
\\
& \times
\cos^{2} \left[ 
\frac{q^{2} e^{- 2 \omega}}{2 \pi }
\left[
\left(
\frac{p^{\prime}}{\omega_{p^{\prime}}}
\right)^{2}
+
\left(
\frac{p}{\omega_{p}}
\right)^{2}
\right]
-
\frac{\pi }{4} \right]
\, ,
\label{probability-bar-u-k-u-case-1-num-infty}
\\
\vert \mathcal{M}_{0, -}  (\chi ) \vert^{2}
& \simeq
\frac{\pi q^{2}}{\omega_{p}^{2}} 
\left(
	\omega_{p^{\prime}} \omega_{p} -   M^{2}
\right)
\frac{e^{-2 \omega}}{\omega^{2}}
\left[
\frac{q^{2} e^{- 2 \omega}}{4 \pi }
\left[
\left(
\frac{p^{\prime}}{\omega_{p^{\prime}}}
\right)^{2}
+
\left(
\frac{p}{\omega_{p}}
\right)^{2}
\right]
\right]^{-1}
\nonumber
\\
& \times
\cos^{2} \left[ 
\frac{q^{2} e^{- 2 \omega}}{2 \pi }
\left[
\left(
\frac{p^{\prime}}{\omega_{p^{\prime}}}
\right)^{2}
+
\left(
\frac{p}{\omega_{p}}
\right)^{2}
\right]
-
\frac{3 \pi }{4} \right]
\, ,
\label{probability-bar-u-epsilon-p-u-case-1-num-infty}
\\
\vert \mathcal{M}_{0, +}  (\chi ) \vert^{2}
& \simeq
\frac{\pi q^{2}}{\omega_{p}^{2}} 
\left(
	\omega_{p^{\prime}} \omega_{p} -   M^{2}
\right)
\frac{e^{-2 \omega}}{\omega^{2}}
\left[
\frac{q^{2} e^{- 2 \omega}}{4 \pi }
\left[
\left(
\frac{p^{\prime}}{\omega_{p^{\prime}}}
\right)^{2}
+
\left(
\frac{p}{\omega_{p}}
\right)^{2}
\right]
\right]^{-1}
\nonumber
\\
& \times
\cos^{2} \left[ 
\frac{q^{2} e^{- 2 \omega}}{2 \pi }
\left[
\left(
\frac{p^{\prime}}{\omega_{p^{\prime}}}
\right)^{2}
+
\left(
\frac{p}{\omega_{p}}
\right)^{2}
\right]
+
\frac{\pi }{4} \right]
\, .
\label{probability-bar-u-epsilon-m-u-case-1-num-infty}
\end{align}
The above relations show that the probability amplitudes of the partial processes can be understood as functions that oscillate with cosine squared (actually a superposition of sine and cosine in the last equations due to the corresponding phases) but with amplitudes damped as either $\omega^{-2 \omega}$ or $\omega^{-4}$ with the exception of $\vert \mathcal{M}_{0, \varepsilon^{\prime}}  (\chi ) \vert^{2}$ which increases as $e^{2 \omega}$. We note that the oscillating nature of these amplitudes is similar to the one obtained in the case of laser wave backgrounds (see e. g. \cite{Seipt:2010ya,Seipt:2016,Mackenroth:2012rb}).

\subsubsection{Case $m = 1, s = 0$, $n = 0$}

It is interesting to consider a case where we fix the knot parameters to $(m = 1, s = 0)$ but keep most of the other parameter as close as possible to the previous case. In this case, we consider the partial amplitudes for the following states
\begin{align}
k_{\mu}	&	=	\left( k_{0} , k_{1}, k_{2}, k_{3} \right) 
			=	\left( \omega , k_{1} = k, 0, 0 \right)
\, ,
\nonumber 
\\
\varepsilon^{\mu}_{1} & = \left( 0, \boldsymbol{\varepsilon}_{1} \right)
			=	\left( 0, k_{1} k_{3}, k_{2} k_{3}, 
			- k^{2}_{1} - k^{2}_{2} \right)
\nonumber
\\
\varepsilon^{\mu}_{2} & = \left( 0, \boldsymbol{\varepsilon}_{2} \right)
			=	\left( 0, - \omega k_{2}, \omega k_{1}, 0 \right)
\nonumber
\\  
p_{\mu}	&	=	\left( p_{0} , p_{1}, p_{2}, p_{3} \right) 
			=	\left( \omega_{p} , 0, 0, p_{3} = p \right)
\, ,
\nonumber
\\
p^{\prime}_{\mu}	&	=	\left( p^{\prime}_{0} , p^{\prime}_{1}, 
							p^{\prime}_{2}, p^{\prime}_{3}
						\right) 
					=	\left( \omega_{p^{\prime}}, 0, 0 , 
						p^{\prime}_{3} = p^{\prime} 
						\right)
\, ,
\nonumber
\\
\varepsilon^{\star \prime}_{\lambda^{\prime}, 1, \mu} & = 
	\left( 
	\varepsilon^{\star \prime \mu}_{\lambda^{\prime}, 1, 0}, 
	\boldsymbol{\varepsilon}^{\star \prime}_{\lambda^{\prime} , 1} 
	\right)
	= \left( 0, 0, 0, 1 \right)
\nonumber
\\
\varepsilon^{\star \prime}_{\lambda^{\prime}, 2, \mu} & = 
	\left( 
	\varepsilon^{\star \prime \mu}_{\lambda^{\prime}, 2, 0}, 
	\boldsymbol{\varepsilon}^{\star \prime}_{\lambda^{\prime} , 2} 
	\right)
	= \left( 0, 0, 0, 0 \right)
\, ,				
\label{fixed-states-Case-2}
\end{align}
This choice of particle and photon states is instructive since it exemplifies  
the symmetry between the set of states and the associated particle amplitudes that depend on the knot numbers $(m,s)$ of the wave backgrounds. 

Indeed, by repeating the same steps as in the previous case, we obtain the following amplitudes
\begin{align} 
\vert \mathcal{M}_{0, \varepsilon^{\prime}}  (\chi ) \vert^{2}
& =
8 \pi^{2} 
\left(
	 \omega_{p^{\prime}} \omega_{p} 
	 +
	 p^{\prime} p - M^{2} 
\right)
\vert
J_{0} \left( \sqrt{\Gamma^{2}_{1} + \Gamma^{2}_{2}} \right)
\vert^{2}
\, ,
\label{probability-bar-u-epsilon-u-Case-2}
\\
\vert \mathcal{M}_{0, k}  (\chi ) \vert^{2} 
& =
\frac{\pi^{2} q^{2} \omega_{p^{\prime}} }{2 \omega_{p} } 
\frac{e^{-2 \omega}}{\omega^{4}}
\vert
J_{0} \left( \sqrt{\Gamma^{2}_{1} + \Gamma^{2}_{2}} \right)
\vert^{2}
\, ,
\label{probability-bar-u-k-u-Case-2}
\\
\vert \mathcal{M}_{0, \pm}  (\chi ) \vert^{2}
& =
\frac{\pi^{2} q^{2} }{2  \omega_{p}^{2}} 
\left( 
\omega_{p^{\prime}} \omega_{p}  -  M^{2}
\right)
\frac{e^{-2 \omega}}{\omega^{2}}
\vert
J_{\mp 1} \left( \sqrt{\Gamma^{2}_{1} + \Gamma^{2}_{2}} \right)
\vert^{2}
\, .
\label{probability-bar-u-epsilon-pm-u-Case-2}
\end{align}
Note that these amplitudes are identical to the ones from the previous case given by the equations 
(\refeq{probability-bar-u-epsilon-u-case-1}),
(\refeq{probability-bar-u-k-u-case-1})
and
(\refeq{probability-bar-u-epsilon-pm-u-case-1}).
Moreover, by calculating the argument of the Bessel function we obtain the following result
\begin{align}
\Gamma^{2}_{1} + \Gamma^{2}_{2}
& = 
\frac{q^{2} e^{- 2 \omega}}{2 \pi }
\left(
	\frac{ p^{\prime}}{\omega_{p^{\prime}}}
	-
	\frac{  p}{\omega_{p}}
	\right)^{2}
\, ,
\label{Gamma-1-Gamma-2-fin-2-Case-2-fin}
\end{align}
which is the same as the one obtained in the equation 
(\refeq{Gamma-1-Gamma-2-fin-2-particular-1}).
Finally, we obtain for the effective mass the function
\begin{equation}
M^{2}_{\star}  = 
M^{2}
\left(
1 +
\frac{q^{2} e^{-2\omega}}{2 \pi  M^2}
\right)
\, .
\label{effective-mass-final-case-2}
\end{equation}
This is the same as the mass given by the equation (\refeq{effective-mass-final-case-1}).

We conclude that the values of the probabilities of the partial amplitudes, the argument of the Bessel functions and the effective mass are the same in the two case $(m=0, s = 1)$ and $(m=1, s = 0)$ but for different states. Note that this result is not trivial since none of these function is symmetric in $m$ and $s$.
Therefore, it is meaningful to ask whether the coincidence of these expressions is an accident or is a result of a deeper relationship between the states and a consequence of the geometrical properties of the knot wave background which inherits symmetry properties from the topological knot.
From the practical side, since the computation of probabilities for more general knot numbers $(m,s)$ is obviously more difficult, the question about the existence of a symmetry which could help simplify these computations is relevant. 

Due to the identical results obtained in the case $(m = 1, s = 0)$, it follows that the discussion of the limits $\zeta \to 0$ and $\zeta \to \infty$ of the probability functions are the same as in the case $m=0$, $s=1$.

\section{Conclusions}

In this paper, we have analysed the dynamics of a charged particle in the background of a strong electromagnetic non-null knot wave field at classical as well as quantum level. The interest in this type of system is motivated by the recent theoretical studies of the properties of the electromagnetic knots as well as by the experimental results that have reported the generation of electromagnetic phase knots of which the knots fields discussed in this paper represent a generalization. 

From the point of view of the investigation of the topological solutions of Maxwell's equations, the results obtained here provide new information on the physics of systems with topological electromagnetic fields. 
Here, we have considered the solutions parametrized by the integers with $n = m $ and $l = s$ which describe a two-parameter family of knots for which the solutions of the quantum equations of motion are superpositions of the solutions of two independent waves.
At the classical level, we have completely determined the particle dynamics in the Hamilton-Jacobi formalism and have shown that our results can be used to calculate relevant physical quantities such as the energy scattered per unit frequency and solid angle. The quantum properties have been discussed in the case of the spin half particle. We have argued that for wave frequencies much higher than a critical frequency $\omega_{cr}$ that was estimated in the equation (\refeq{estimate-strong-frequencies}), the knot wave can be treated as a classical strong background field. We have quantized this system by using canonical methods of the strong-field QED. In particular, we have constructed the Furry picture and have given the Fock space and the operatorial equations of motion for the Dirac field and the radiation field in the knot background. Then, by using the strong-QED techniques found in the literature in standard texts, e. g. 
\cite{Fradkin:1981sc,Furry:1951zz,Mackenroth:2014}, we have solved the Dirac equation and determined a new set of Volkov solutions in the real monochromatic electromagnetic knot given by the equation (\refeq{Volkov-solution-knot-2}). These Volkov functions form a family paramerized by the knot numbers $(m,s)$ which contain information about the topological structure of the original complex knot wave. However, this information is reduced when compared with the full knot wave by the procedure of taking the real part of the complex wave and keeping
the orthogonal superposition of sine and cosine background fields.
The general results of the strong-field QED give the elements of the $S$-matrix in terms of Volkov fields which allow us to calculate the probabilities of the processes involving charged spin half particles and photons in the real knot wave background. We have exemplified our results 
in the case of the one-photon Compton effect for which we have given the power expansion of the $S$-matrix in the interaction parameter and have determined its form and the average differential emission probability in the lowest order approximation.

The results obtained in this paper open up a series of interesting investigation lines. One interesting problem is to find all mathematical properties of the Ritus matrices and implicitly the Volkov solutions for the knot wave backgrounds. That is an important issue even in the general context of electric charges in strong plane-wave fields, with new mathematical properties proved recently in e. g. \cite{Boca:2011}. The Volkov solutions obtained here can be generalized to wave packets, which is interesting for the theory of laser beams. While the full knot is such of packet, it is certainly interesting to study the properties of wave packets with different momentum distributions or carrier envelope phase. Naturally, it is possible to repeat almost verbatim the present calculations for other electromagnetic knots. 
One important point here is to discuss the regularization of the quantum amplitudes in the presence of the electromagnetic potentials which would allow one to compare the results with the Coulomb scattering in other types of knot backgrounds. As we have seen in the discussion of the wave backgrounds with the knot numbers $(m=0, s= 1)$ and $(m=1, s = 0)$, there is a symmetry parametrized by the $(m,s)$ among the partial amplitudes of different states. It is a meaningful question whether this symmetry is a general property of the $S$-matrix or just a casual property of some states and knot backgrounds. Since the knot waves are solutions of Maxwell's equations in vacuum, the whole quantization procedure its based on the quantization of the fluctuations of the electromagnetic field around these solutions. Therefore, it is necessary to analyse further the stability of the knot backgrounds in the presence of sources. Also, it is important to investigate other quantum effects in the presence of the knot fields, like e. g. the two-photon Compton effect \cite{Seipt:2010ya,Mackenroth:2012rb}, in order to understand better these intriguing systems. And finally, from the classical point of view, it would be interesting to investigate whether the superintegrability methods developed in 
\cite{Heinzl:2017zsr,Heinzl:2017blq} apply to the the knot waves.

\section*{Acknowledgements} I. V. V. would like to thank to J. A. Helay\"{e}l-Neto for hospitality at CBPF where part of this work was done and to H. Nastase for correspondence on electromagnetic knots. Also, we acknowledge important suggestions and comments of an anonymous referee that helped us to improve the content of the text.

\section*{Appendix A: Fourier decomposition of knot fields}
\renewcommand{\theequation}{A.\arabic{equation}}
\setcounter{equation}{0}

Here, we gives some more details on the Fourier decomposition of the non-null knot fields discussed thoroughly in \cite{Arrayas:2018}.

By applying the Fourier transform to the electromagnetic field in vacuum in the Lorentz gauge, one obtains the following expressions
\begin{align}
\mathbf{E} (x^{0} , \mathbf{x} )
& =
\frac{1}{2 (2 \pi)^{3/2}}
\int d^{3} k
\left[
\left(
\mathbf{E}^{\star}_{0} (\mathbf{k}) - \mathbf{n} \times 
\mathbf{B}^{\star}_{0} (\mathbf{k})
\right) e^{-i k \cdot x}
+ \left(
\mathbf{E}_{0} (\mathbf{k}) - \mathbf{n} \times 
\mathbf{B}_{0} (\mathbf{k})
\right) e^{i k \cdot x}
\right]
\, ,
\label{E-field-Fourier-App-A}
\\
\mathbf{B} (x^{0} , \mathbf{x} )
& =
\frac{1}{2 (2 \pi)^{3/2}}
\int d^{3} k
\left[
\left(
\mathbf{B}^{\star}_{0} (\mathbf{k}) + \mathbf{n} \times 
\mathbf{E}^{\star}_{0} (\mathbf{k})
\right) e^{-i k \cdot x}
+ \left(
\mathbf{B}_{0} (\mathbf{k}) + \mathbf{n} \times 
\mathbf{E}_{0} (\mathbf{k})
\right) e^{i k \cdot x}
\right]
\, .
\label{B-field-Fourier-App-A}
\end{align} 
From their definitions, it follows that the corresponding three dimensional potentials $\mathbf{A}$ and $\mathbf{C}$ have the Fourier decompositions
\begin{align}
\mathbf{A} (x^{0} , \mathbf{x} )
& =
\frac{i}{ (2 \pi)^{3/2}}
\int d^{3} k
\left[
\left(
- \mathbf{E}^{\star}_{0} (\mathbf{k}) + \mathbf{n} \times 
\mathbf{B}^{\star}_{0} (\mathbf{k})
\right) e^{-i k \cdot x}
+ \left(
\mathbf{E}_{0} (\mathbf{k}) - \mathbf{n} \times 
\mathbf{B}_{0} (\mathbf{k})
\right) e^{i k \cdot x}
\right]
\, ,
\label{A-field-Fourier-App-A}
\\
\mathbf{C} (x^{0} , \mathbf{x} )
& =
\frac{i}{ (2 \pi)^{3/2}}
\int d^{3} k
\left[
\left(
\mathbf{B}^{\star}_{0} (\mathbf{k}) + \mathbf{n} \times 
\mathbf{E}^{\star}_{0} (\mathbf{k})
\right) e^{-i k \cdot x}
+ \left(
- \mathbf{B}_{0} (\mathbf{k}) + \mathbf{n} \times 
\mathbf{E}_{0} (\mathbf{k})
\right) e^{i k \cdot x}
\right]
\, .
\label{C-field-Fourier}
\end{align} 
The potentials can be written in terms of circularly polarized waves given by the relations (\refeq{epsilon-plus-App-A}) and (\refeq{epsilon-minus-App-A}) below  upon the identification
\begin{align}
a_{R} (\mathbf{k}) \, \mathbf{e}_{R}(\mathbf{k})
+ a_{L} (\mathbf{k}) \, \mathbf{e}_{L} (\mathbf{k}) & = 
\frac{-i}{\sqrt{2 k}}
\left(
\mathbf{E}^{\star}_{0} (\mathbf{k}) - \mathbf{n} \times 
\mathbf{B}^{\star}_{0} (\mathbf{k})
\right)
\, ,
\label{epsilon-plus-E-B-App-A}
\\
a_{R} (\mathbf{k}) \, \mathbf{e}_{R}(\mathbf{k})
- a_{L} (\mathbf{k}) \, \mathbf{e}_{L} (\mathbf{k}) & = 
\frac{1}{\sqrt{2 k}}
\left(
\mathbf{B}^{\star}_{0} (\mathbf{k}) + \mathbf{n} \times 
\mathbf{E}^{\star}_{0} (\mathbf{k})
\right)
\, .
\label{epsilon-minus-E-B-App-A}
\end{align}
The vectors of the basis satisfy the following product relations 
\begin{align}
\mathbf{e}_R \cdot \mathbf{e}_R & = 0
\, ,
\quad
\mathbf{e}_L \cdot \mathbf{e}_L = 0
\, ,
\quad
\mathbf{n} \cdot \mathbf{n}  = 1
\, ,
\nonumber
\\ 
\quad
\mathbf{e}_L \times \mathbf{e}_R & = i \mathbf{n}
\, ,
\quad
\mathbf{e}_R \times \mathbf{n}  = i \mathbf{e}_R
\, ,
\quad
\mathbf{n} \times \mathbf{e}_L = i \mathbf{e}_L 
\, .
\label{helicity-basis-properties-App-A}
\end{align}
Here, $ \mathbf{e}_{R} (\mathbf{k})$ and $\mathbf{e}_{L} (\mathbf{k})$ are the left and right linearly independent helicity vectors and $a_{R} (\mathbf{k})$ and $a_{L} (\mathbf{k})$ are the Fourier coefficients in the helicity basis which is defined by the set of three-dimensional complex vectors $\{ \mathbf{e}_R (\mathbf{k}) , \mathbf{e}_L (\mathbf{k}), \mathbf{n} (\mathbf{k}) = \mathbf{k}/ \vert \mathbf{k} \vert \}$.  
By applying the above relations to the knots defined by the equations (\refeq{E-non-null-knot}) and (\refeq{B-non-null-knot}), one obtains the Fourier decomposition of the corresponding potentials. In general, the Fourier modes are complex vector fields \cite{Arrayas:2018} of the following form
\begin{align}
\mathbf{A} (x^{0}, \mathbf{x}) & =
\int d \mu (\mathbf{k} ) 
\left[ \boldsymbol{\epsilon}^{\mu}_{+} (\mathbf{k}) e^{-i k \cdot x}
+ \mbox{c. c.}
\right] 
\, ,
\label{A-Fourier-original-App-A}
\\
\mathbf{C} (x^{0}, \mathbf{x}) & =
\int d \mu (\mathbf{k} ) 
\left[ \boldsymbol{\epsilon}^{\mu}_{-} (\mathbf{k}) e^{-i k \cdot x}
+ \mbox{c. c.}
\right] 
\, .
\label{C-Fourier-original-App-A}
\end{align}
Here, we are using the following notation for the covariant measure in the particle phase space and photon phase space, respectively,
\begin{equation}
d \mu(\mathbf{p}) = \frac{d^{3} \mathbf{p}}{\left( 2 \pi \right)^{3} 2 \omega_{p}}
\, ,
\quad
d \mu(\mathbf{k}) = \frac{d^{3} \mathbf{k}}{\left( 2 \pi \right)^{3} 2 \omega_{k}}
\, .
\label{covariant-measures-App-A}
\end{equation}
One can easily see that the three-dimensional polarization vectors can be written as follows
\begin{align}
\boldsymbol{\epsilon}^{\mu}_{+} \left( \mathbf{k} \right) & = 
a_{R} (\mathbf{k}) \, \mathbf{e}_{R}(\mathbf{k})
+ a_{L} (\mathbf{k}) \, \mathbf{e}_{L} (\mathbf{k})
\, ,
\label{epsilon-plus-App-A}
\\
\boldsymbol{\epsilon}^{\mu}_{-} \left( \mathbf{k} \right) & = 
a_{R} (\mathbf{k}) \, \mathbf{e}_{R}(\mathbf{k})
- a_{L} (\mathbf{k}) \, \mathbf{e}_{L} (\mathbf{k})
\, .
\label{epsilon-minus-App-A}
\end{align}
It follows that the explicit form of the products from the right hand side of the equations (\refeq{epsilon-plus-App-A}) and (\refeq{epsilon-minus-App-A}) in the basis $\{ \mathbf{e}_{1}, \mathbf{e}_{2}, \mathbf{e}_3 \}$ and in the $\mathbf{k}$-space are
\begin{align}
a_{L} (\mathbf{k}) \, \mathbf{e}_{L} (\mathbf{k}) & =
\frac{1}{4 \sqrt{\pi \vert \mathbf{k} \vert }} e^{- \vert \mathbf{k} \vert}
\left\{
\left\{
\frac{m - n}{\vert \mathbf{k} \vert}
\begin{bmatrix}
	k_1 k_3 
	\\
	k_2 k_3 
	\\
	- \left( k^{2}_{1} + k^{2}_{2} \right)
\end{bmatrix}
+
\left( s - l \right)
\begin{bmatrix}
	0
	\\
	k_3
	\\
	- k_2
\end{bmatrix}
\right\}
\right.
\nonumber
\\
& - i
\left.
\left\{
\frac{l - s}{\vert \mathbf{k} \vert}
\begin{bmatrix}
	k^{2}_{2} + k^{2}_{3} 
	\\
	- k_1 k_2 
	\\
	- k_1 k_3
\end{bmatrix}
+
\left( n - m \right)
\begin{bmatrix}
	k_2
	\\
	- k_1
	\\
	0
\end{bmatrix}
\right\}
\right\}
\, ,
\label{aLeL-App-A}
\\
a_{R} (\mathbf{k}) \, \mathbf{e}_{R} (\mathbf{k}) & =
\frac{1}{4 \sqrt{\pi \vert \mathbf{k} \vert }} e^{- \vert \mathbf{k} \vert}
\left\{
\left\{
\frac{ m + n}{\vert \mathbf{k} \vert}
\begin{bmatrix}
	k_1 k_3 
	\\
	k_2 k_3 
	\\
	- \left( k^{2}_{1} + k^{2}_{2} \right)
\end{bmatrix}
+
\left( s + l \right)
\begin{bmatrix}
	0
	\\
	k_3
	\\
	- k_2
\end{bmatrix}
\right\}
\right.
\nonumber
\\
& - i
\left.
\left\{
\frac{s + l}{\vert \mathbf{k} \vert}
\begin{bmatrix}
	k^{2}_{2} + k^{2}_{3} 
	\\
	- k_1 k_2 
	\\
	- k_1 k_3
\end{bmatrix}
+
\left( n + m \right)
\begin{bmatrix}
	k_2
	\\
	- k_1
	\\
	0
\end{bmatrix}
\right\}
\right\}
\, ,
\label{aReR-App-A}
\end{align}
where $\mathbf{k} = (k_1 , k_2, k_3)$. The normalization of the vectors of the Cartesian basis and the relationship between the two basis are 
\begin{align}
\mathbf{e}_{1}^{2} &= \mathbf{e}_{2}^{2} = 1/2 
\, ,
\quad \mathbf{e}_3^{2} = 1
\, ,
\label{basis-1-App-A}
\\
\mathbf{e}_R & = \mathbf{e}_{1} + i \mathbf{e}_{2}
\, ,
\quad 
\mathbf{e}_L = \mathbf{e}_{1} - i \mathbf{e}_{2}
\, , 
\quad \mathbf{n} = \mathbf{e}_3
\, .
\label{basis-2-App-A}
\end{align}

\section*{Appendix B: Some partial amplitudes in Compton scattering}
\renewcommand{\theequation}{B.\arabic{equation}}
\setcounter{equation}{0}

Here, we present the calculation of the amplitude $\vert \mathcal{M}_{n, \varepsilon^{\prime}} \mathcal{M}^{\star}_{n, \mp} \vert$ in terms of the knot parameters $(m,s)$ in the case of arbitrary momenta.

By using their definitions, we obtain the following expressions for the arguments of the Bessel and the exponential functions
\begin{align}
\frac{\Gamma_{1}}{\Gamma_{2}}
& = 
\left\{
	\left[
	- m k_{1} k_{2} p^{\prime}_{1} + 
		\left(
		m k_{1} k_{3} + s \omega k_{3}
		\right) p^{\prime}_{2}
		+
		\left[
		m 
		\left(
		k^{2}_{1} + k^{2}_{2}
		\right)
		+ s \omega k_{2}
		\right] p^{\prime}_{3}
	\right]
	\left(
	p \cdot k
	\right)
\right.
\nonumber
\\
& -
\left.
	\left[
	- m k_{1} k_{2} p_{1} + 
		\left(
		m k_{1} k_{3} + s \omega k_{3}
		\right) p_{2}
		+
		\left[
		m 
		\left(
		k^{2}_{1} + k^{2}_{2}
		\right)
		+ s \omega k_{2}
		\right] p_{3}
	\right]
	\left(
	p^{\prime} \cdot k
	\right)
\right\}
\nonumber
\\
& \times
\left\{
	\left[
	\left[ 
		m \omega k_{2} 
		+ s	\left(
		k^{2}_{2} + k^{2}_{3}
		\right) 
	\right] p^{\prime}_{1}
	-
	\left(
		m \omega k_{1} + s k_{1} k_{2}
	\right) p^{\prime}_{2}
	-
	s k_{2}k_{3} p^{\prime}_{3}
	\right]
	\left(
	p \cdot k
	\right)	
\right.
\nonumber
\\
& -
	\left[
	\left[ 
		m \omega k_{2} 
		+ s	\left(
		k^{2}_{2} + k^{2}_{3}
		\right) 
	\right] p_{1}
	-
	\left(
		m \omega k_{1} + s k_{1} k_{2}
	\right) p_{2}
	-
	s k_{2}k_{3} p_{3}
	\right]
	\left(
	p^{\prime} \cdot k
	\right)	
\left.
\right\}^{-1}
\, ,
\label{Gamma-1-Gamma-2-fin-1-App-B}
\\
\Gamma^{2}_{1} + \Gamma^{2}_{2}
& =
\frac{q^{2} e^{- 2 \omega}}{2 \pi \omega^{2}}
\left\{
\left\{
	\left[
	- m k_{1} k_{2} p^{\prime}_{1} + 
		\left(
		m k_{1} k_{3} + s \omega k_{3}
		\right) p^{\prime}_{2}
	+	\left[
		m
			\left(
			k^{2}_{1} + k^{2}_{2}
			\right)
		+ s \omega k_{2}
		\right] p^{\prime}_{3}
	\right]^{2}
\right.
\right.
\nonumber
\\
& -
\left.
	\left[
	\left[
	m \omega k_{2} 
	+ 
	s	\left(
		k^{2}_{2} + k^{2}_{3}
		\right)
	\right] p^{\prime}_{1}
	-
	\left(
		m \omega k_{1} + s k_{1} k_{2}
	\right) p^{\prime}_{2}
	- s k_{2}k_{3} p^{\prime}_{3}
	\right]^{2}
\right\}
\times
	\left( 
	p^{\prime} \cdot k
	\right)^{-2}
\nonumber
\\
& +
\left\{
	\left[
	- m k_{1} k_{2} p_{1} + 
		\left(
		m k_{1} k_{3} + s \omega k_{3}
		\right) p_{2}
	+	\left[
		m
			\left(
			k^{2}_{1} + k^{2}_{2}
			\right)
		+ s \omega k_{2}
		\right] p_{3}
	\right]^{2}
\right.
\nonumber
\\
& -
\left.
	\left[
	\left[
	m \omega k_{2} 
	+ 
	s	\left(
		k^{2}_{2} + k^{2}_{3}
		\right)
	\right] p_{1}
	-
	\left(
		m \omega k_{1} + s k_{1} k_{2}
	\right) p_{2}
	- s k_{2}k_{3} p_{3}
	\right]^{2}
\right\}
\times
	\left( 
	p \cdot k
	\right)^{-2}
\nonumber
\\
& - 2
\left\{
	\left[
	- m k_{1} k_{2} p^{\prime}_{1} + 
	\left( 
		m k_{1} k_{3} + s \omega k_{3} 
	\right) p^{\prime}_{2}
	+	\left[
			m 
			\left(
				k^{2}_{1} + k^{2}_{2}
			\right)
			+ s \omega k_{2}
		\right] p^{\prime}_{3}
	\right]
\right.
\nonumber
\\
& \times
	\left[
	- m k_{1} k_{2} p_{1} + 
	\left( 
		m k_{1} k_{3} + s \omega k_{3} 
	\right) p_{2}
	+	\left[
			m 
			\left(
				k^{2}_{1} + k^{2}_{2}
			\right)
			+ s \omega k_{2}
		\right] p_{3}
	\right]
\nonumber
\\
& +
	\left[
	\left[
	m \omega k_{2} 
	+ s	\left(
			k^{2}_{2} + k^{2}_{3}
		\right) 
	\right] p^{\prime}_{1}
	- 
	\left(
	m \omega k_{1} + s k_{1} k_{2}
	\right) p^{\prime}_{2}
	- s k_{2}k_{3} p^{\prime}_{3}
	\right]
\nonumber
\\
& \times
\left.
\left.
	\left[
	\left[
	m \omega k_{2} 
	+ s	\left(
			k^{2}_{2} + k^{2}_{3}
		\right) 
	\right] p_{1}
	- 
	\left(
	m \omega k_{1} + s k_{1} k_{2}
	\right) p_{2}
	- s k_{2}k_{3} p_{3}
	\right]
\right\}
	\left(
	p^{\prime} \cdot k
	\right)^{-1}
	\left(
	p \cdot k
	\right)^{-1}
\right\}
\label{Gamma-1-Gamma-2-fin-2-App-B}
\end{align}
Since these are quite large expressions, they have not been substituted in the formulas where they appear in the text. 

The real partial amplitude that we calculate in terms of $(m,s)$ has the following form
\begin{equation}
\Re \left[ \mathcal{M}_{n, \varepsilon^{\prime}}  (\chi ) 
\mathcal{M}^{\star}_{n, \mp}  (\chi ) \right] = 
\Re \left[ \mathcal{M}_{n, \varepsilon^{\prime}}  (\chi ) 
\mathcal{M}^{\star}_{n, \mp}  (\chi ) \right]_{1}
+
\Re \left[ \mathcal{M}_{n, \varepsilon^{\prime}}  (\chi ) 
\mathcal{M}^{\star}_{n, \mp}  (\chi ) \right]_{2}
\, .
\label{real-part-M-M-general-App}
\end{equation}
After some extended but straightforward calculations, the relation (\refeq{real-part-M-M-general-App}) takes the following form
\begin{align}
&\Re \left[ \mathcal{M}_{n, \varepsilon^{\prime}}  (\chi ) 
\mathcal{M}^{\star}_{n, \mp}  (\chi ) \right]
=
\nonumber
\\
& - \frac{4 \pi \sqrt{2 \pi} q e^{- \omega}}{\omega^{2}p \cdot k} 
\left\{
% FIRST BLOCK %%%%%%%%%%%%%%%%%%%%%%%%%%%%%%%%%%%%%%%%%
	\left(
	k \cdot \varepsilon^{\star \prime}_{\lambda^{\prime}, 1}
	\right)
	\left[
	- m k_{1} k_{2} p^{\prime}_{1} + 
	\left(
		m k_{1} k_{3} + s \omega k_{3}
	\right) p^{\prime}_{2}
	+
	\left[
	m 
		\left(
		k^{2}_{1} + k^{2}_{2}
		\right)
	+ s \omega k_{2}
	\right] p^{\prime}_{3}
	\right]
\right.
\nonumber
\\
& \times
	\left[
		\left(
		p \cdot \varepsilon^{\star \prime}_{\lambda^{\prime}, 1}
		\right)
		\sin \left(\arctan\left(\frac{\Gamma_{1}}{\Gamma_{2}}\right) \right)
		\mp
		\left(
		p \cdot \varepsilon^{\star \prime}_{\lambda^{\prime}, 2}
		\right)
		\cos \left(\arctan\left(\frac{\Gamma_{1}}{\Gamma_{2}}\right) \right)
	\right]
\nonumber
\\
& +
	\left(
	k \cdot \varepsilon^{\star \prime}_{\lambda^{\prime}, 1}
	\right)
	\left[
	- m k_{1} k_{2} p_{1} + 
	\left(
		m k_{1} k_{3} + s \omega k_{3}
	\right) p_{2}
	+
	\left[
	m 
		\left(
		k^{2}_{1} + k^{2}_{2}
		\right)
	+ s \omega k_{2}
	\right] p_{3}
	\right]
\nonumber
\\
& \times
	\left[
		\left(
		p^{\prime} \cdot \varepsilon^{\star \prime}_{\lambda^{\prime}, 1}
		\right)
		\sin \left(\arctan\left(\frac{\Gamma_{1}}{\Gamma_{2}}\right) \right)
		\mp
		\left(
		p^{\prime} \cdot \varepsilon^{\star \prime}_{\lambda^{\prime}, 2}
		\right)
		\cos \left(\arctan\left(\frac{\Gamma_{1}}{\Gamma_{2}}\right) \right)
	\right]
\nonumber
\\
& -
	\left(
	k \cdot \varepsilon^{\star \prime}_{\lambda^{\prime}, 1}
	\right)
	\left(
	p \cdot p^{\prime} - M^{2}
	\right)
\nonumber
\\
& \times
	\left[
	\left[
	- m k_{1} k_{2} 
	\varepsilon^{\star \, \prime}_{\lambda^{\prime}, 1, 1}
	+ 
	\left(
		m k_{1} k_{3} + s \omega k_{3}
	\right) 
	\varepsilon^{\star \, \prime}_{\lambda^{\prime}, 1, 2}
	+
	\left[
		m 
			\left(
				k^{2}_{1} + k^{2}_{2}
			\right)
		+ s \omega k_{2}
	\right] 
	\varepsilon^{\star \, \prime}_{\lambda^{\prime}, 1, 3}
	\right]
	\sin \left(\arctan\left(\frac{\Gamma_{1}}{\Gamma_{2}}\right) \right)
	\right.
\nonumber
\\
& \mp
	\left.
	\left[
	\left[ 
	m \omega k_{2} 
	+ s	\left(
		k^{2}_{2} + k^{2}_{3}
		\right) 
	\right]
	\boldsymbol{\varepsilon}^{\star \, \prime}_{\lambda^{\prime}, 2, 1}
	-
		\left(
		m \omega k_{1} + s k_{1} k_{2}
		\right) 
	\boldsymbol{\varepsilon}^{\star \, \prime}_{\lambda^{\prime}, 2, 2}
	- s k_{2}k_{3} 
	\boldsymbol{\varepsilon}^{\star \, \prime}_{\lambda^{\prime}, 2, 3}
	\right]
	\cos \left(\arctan\left(\frac{\Gamma_{1}}{\Gamma_{2}}\right) \right)
	\right]
\nonumber
\\
% SECOND BLOCK %%%%%%%%%%%%%%%%%%%%%%%%%%%%%%%%%%%%%%%%%%%%%
& \pm
	\left(
	k \cdot \varepsilon^{\star \prime}_{\lambda^{\prime}, 2}
	\right)
	\left[
	\left[ m \omega k_{2} 
	+ s	\left(
		k^{2}_{2} + k^{2}_{3}
		\right) 
	\right] p^{\prime}_{1}
	- \left( 
	m \omega k_{1} + s k_{1} k_{2} 
	\right) p^{\prime}_{2} 
	- s 
	k_{2}k_{3} p^{\prime}_{3}
	\right]
\nonumber
\\
& \times
	\left[
		\left(
		p \cdot \varepsilon^{\star \prime}_{\lambda^{\prime}, 1}
		\right)
		\sin \left(\arctan\left(\frac{\Gamma_{1}}{\Gamma_{2}}\right) \right)
		\mp
		\left(
		p \cdot \varepsilon^{\star \prime}_{\lambda^{\prime}, 2}
		\right)
		\cos \left(\arctan\left(\frac{\Gamma_{1}}{\Gamma_{2}}\right) \right)
	\right]
\nonumber
\\
& \pm
	\left(
	k \cdot \varepsilon^{\star \prime}_{\lambda^{\prime}, 2}
	\right)
	\left[
	\left[ m \omega k_{2} 
	+ s	\left(
		k^{2}_{2} + k^{2}_{3}
		\right) 
	\right] p_{1}
	- \left( 
	m \omega k_{1} + s k_{1} k_{2} 
	\right) p_{2} 
	- s 
	k_{2}k_{3} p_{3}
	\right]
\nonumber
\\
& \times
	\left[
		\left(
		p^{\prime} \cdot \varepsilon^{\star \prime}_{\lambda^{\prime}, 1}
		\right)
		\sin \left(\arctan\left(\frac{\Gamma_{1}}{\Gamma_{2}}\right) \right)
		\mp
		\left(
		p^{\prime} \cdot \varepsilon^{\star \prime}_{\lambda^{\prime}, 2}
		\right)
		\cos \left(\arctan\left(\frac{\Gamma_{1}}{\Gamma_{2}}\right) \right)
	\right]
\nonumber
\\
& \mp
	\left(
	k \cdot \varepsilon^{\star \prime}_{\lambda^{\prime}, 2}
	\right)
	\left(
	p \cdot p^{\prime} - M^{2}
	\right)
\nonumber
\\
& \times
	\left[
	\left[
		\left[
			m \omega k_{2} 
			+ 
			s	\left(
				k^{2}_{2} + k^{2}_{3}
				\right) 
		\right] 
		\varepsilon^{\star \prime}_{\lambda^{\prime}, 1, 1}
		-
		\left(
			m \omega k_{1} + s k_{1} k_{2}
		\right) 
		\varepsilon^{\star \prime}_{\lambda^{\prime}, 1, 2}
		-
		s k_{2}k_{3} 
		\varepsilon^{\star \prime}_{\lambda^{\prime}, 1, 3}
	\right]
	\sin \left(\arctan\left(\frac{\Gamma_{1}}{\Gamma_{2}}\right) \right)
	\right.
\nonumber
\\
& \mp
	\left.
	\left[
		\left[
			m \omega k_{2} 
			+ 
			s	\left(
				k^{2}_{2} + k^{2}_{3}
				\right) 
		\right] 
		\varepsilon^{\star \prime}_{\lambda^{\prime}, 2, 1}
		-
		\left(
			m \omega k_{1} + s k_{1} k_{2}
		\right) 
		\varepsilon^{\star \prime}_{\lambda^{\prime}, 2, 2}
		-
		s k_{2}k_{3} 
		\varepsilon^{\star \prime}_{\lambda^{\prime}, 2, 3}
	\right]
	\cos \left(\arctan\left(\frac{\Gamma_{1}}{\Gamma_{2}}\right) \right)
	\right]
\nonumber
\\
% THIRD BLOCK %%%%%%%%%%%%%%%%%%%%%%%%%%%%%%%%%%%%%%%
& +
	\left(
	k \cdot \varepsilon^{\star \prime}_{\lambda^{\prime}, 2}
	\right)
	\left[ 
	- m k_{1} k_{2} p^{\prime}_{1} 
	+ 
		\left( 
		m k_{1} k_{3} + s \omega k_{3} 
		\right) p^{\prime}_{2}
	+ 
	\left[ m 
		\left( 
		k^{2}_{1} + k^{2}_{2}
		\right)
	+ s \omega k_{2}
	\right] p^{\prime}_{3}
	\right]
\nonumber
\\
& \times
	\left[
	p \cdot \varepsilon^{\star \prime}_{\lambda^{\prime}, 2}
	\sin \left(\arctan\left(\frac{\Gamma_{1}}{\Gamma_{2}}\right) \right)
	\pm
	p \cdot \varepsilon^{\star \prime}_{\lambda^{\prime}, 1}
	\cos \left(\arctan\left(\frac{\Gamma_{1}}{\Gamma_{2}}\right) \right)
	\right]
\nonumber
\\
& +
	\left(
	k \cdot \varepsilon^{\star \prime}_{\lambda^{\prime}, 2}
	\right)
	\left[ 
	- m k_{1} k_{2} p_{1} 
	+ 
		\left( 
		m k_{1} k_{3} + s \omega k_{3} 
		\right) p_{2}
	+ 
	\left[ m 
		\left( 
		k^{2}_{1} + k^{2}_{2}
		\right)
	+ s \omega k_{2}
	\right] p_{3}
	\right]
\nonumber
\\
& \times
	\left[
	p^{\prime} \cdot \varepsilon^{\star \prime}_{\lambda^{\prime}, 2}
	\sin \left(\arctan\left(\frac{\Gamma_{1}}{\Gamma_{2}}\right) \right)
	\pm
	p^{\prime} \cdot \varepsilon^{\star \prime}_{\lambda^{\prime}, 1}
	\cos \left(\arctan\left(\frac{\Gamma_{1}}{\Gamma_{2}}\right) \right)
	\right]
\nonumber
\\
& -
	\left(
	k \cdot \varepsilon^{\star \prime}_{\lambda^{\prime}, 2}
	\right)
	\left(
	p \cdot p^{\prime} - M^{2}
	\right)
	\left[
	\left[ 
	- m k_{1} k_{2} 
	\varepsilon^{\star \prime}_{\lambda^{\prime}, 2, 1} 
	+	\left(
			m k_{1} k_{3} + s \omega k_{3}
		\right) 
	\varepsilon^{\star \prime}_{\lambda^{\prime}, 2, 2}
	+	\left[ 
			m	\left(
				k^{2}_{1} + k^{2}_{2}
				\right)
		+ s \omega k_{2} 
		\right] 
	\varepsilon^{\star \prime}_{\lambda^{\prime}, 2, 3}
	\right]
	\right.
\nonumber
\\
&\times
	\sin \left(\arctan\left(\frac{\Gamma_{1}}{\Gamma_{2}}\right) \right)
\nonumber
\\
& 
\pm
	\left.
	\left[ 
	- m k_{1} k_{2} 
	\varepsilon^{\star \prime}_{\lambda^{\prime}, 1, 1} 
	+	\left(
			m k_{1} k_{3} + s \omega k_{3}
		\right) 
	\varepsilon^{\star \prime}_{\lambda^{\prime}, 1, 2}
	+	\left[ 
			m	\left(
				k^{2}_{1} + k^{2}_{2}
				\right)
		+ s \omega k_{2} 
		\right] 
	\varepsilon^{\star \prime}_{\lambda^{\prime}, 1, 3}
	\right]
	\cos \left(\arctan\left(\frac{\Gamma_{1}}{\Gamma_{2}}\right) \right)	
	\right]
\nonumber
\\
% FOURTH BLOCK %%%%%%%%%%%%%%%%%%%%%%%%%%%%%%%%%%%%%%%%%%%%%%%%%
& \mp
	\left(
	k \cdot \varepsilon^{\star \prime}_{\lambda^{\prime}, 1}
	\right)
	\left[
	\left[
		m \omega k_{2} 
		+ 
		s	\left(
			k^{2}_{2} + k^{2}_{3}
			\right) 
	\right] p^{\prime}_{1}
	- \left(
			m \omega k_{1} + s k_{1} k_{2}
	\right) p^{\prime}_{2}
	- s 
	k_{2}k_{3} p^{\prime}_{3}
	\right]
\nonumber
\\
& \times
	\left[
	p \cdot \varepsilon^{\star \prime}_{\lambda^{\prime}, 2}
	\sin \left(\arctan\left(\frac{\Gamma_{1}}{\Gamma_{2}}\right) \right)
	\pm
	p \cdot \varepsilon^{\star \prime}_{\lambda^{\prime}, 1}
	\cos \left(\arctan\left(\frac{\Gamma_{1}}{\Gamma_{2}}\right) \right)
	\right]
\nonumber
\\
& \mp
	\left(
	k \cdot \varepsilon^{\star \prime}_{\lambda^{\prime}, 1}
	\right)
	\left[
	\left[
		m \omega k_{2} 
		+ 
		s	\left(
			k^{2}_{2} + k^{2}_{3}
			\right) 
	\right] p_{1}
	- \left(
			m \omega k_{1} + s k_{1} k_{2}
	\right) p_{2}
	- s 
	k_{2}k_{3} p_{3}
	\right]
\nonumber
\\
& \times
	\left[
	p^{\prime} \cdot \varepsilon^{\star \prime}_{\lambda^{\prime}, 2}
	\sin \left(\arctan\left(\frac{\Gamma_{1}}{\Gamma_{2}}\right) \right)
	\pm
	p^{\prime} \cdot \varepsilon^{\star \prime}_{\lambda^{\prime}, 1}
	\cos \left(\arctan\left(\frac{\Gamma_{1}}{\Gamma_{2}}\right) \right)
	\right]
\nonumber
\\
& \pm
	\left(
	k \cdot \varepsilon^{\star \prime}_{\lambda^{\prime}, 1}
	\right)
	\left(
	p \cdot p^{\prime} - M^{2}
	\right)
	\left[
	\left[
	\left[ 
	m \omega k_{2} 
	+ s	\left(
		k^{2}_{2} + k^{2}_{3}
		\right) 
	\right] 
	\varepsilon^{\star \prime}_{\lambda^{\prime}, 2, 1}
	-	\left( 
		m \omega k_{1} + s k_{1} k_{2}
		\right) 
	\varepsilon^{\star \prime}_{\lambda^{\prime}, 2, 2}
	-s k_{2}k_{3} 
	\varepsilon^{\star \prime}_{\lambda^{\prime}, 2, 3}
	\right]
	\right.
\nonumber
\\
&\times
	\sin \left(\arctan\left(\frac{\Gamma_{1}}{\Gamma_{2}}\right) \right)
\nonumber
\\
& \pm
	\left.
	\left[
	\left[ 
	m \omega k_{2} 
	+ s	\left(
		k^{2}_{2} + k^{2}_{3}
		\right) 
	\right] 
	\varepsilon^{\star \prime}_{\lambda^{\prime}, 1, 1}
	-	\left( 
		m \omega k_{1} + s k_{1} k_{2}
		\right) 
	\varepsilon^{\star \prime}_{\lambda^{\prime}, 1, 2}
	-s k_{2}k_{3} 
	\varepsilon^{\star \prime}_{\lambda^{\prime}, 1, 3}
	\right]
	\cos \left(\arctan\left(\frac{\Gamma_{1}}{\Gamma_{2}}\right) \right)
	\right]
\nonumber
\\
% FIFTH BLOCK %%%%%%%%%%%%%%%%%%%%%%%%%%%%%%
& -
	\left[
	\left[
	- m k_{1} k_{2} 
	\varepsilon^{\star \prime}_{\lambda^{\prime}, 1, 1} 
	+	\left(
		m k_{1} k_{3} + s \omega k_{3}
		\right) 
	\varepsilon^{\star \prime}_{\lambda^{\prime}, 1, 2}
	+
	\left[
		m
		\left(
		k^{2}_{1} + k^{2}_{2}
		\right)
		+ s \omega k_{2}
	\right] 
	\varepsilon^{\star \prime}_{\lambda^{\prime}, 1, 3}
	\right]
	\right.
\nonumber
\\
&\pm
	\left.
	\left[
	\left[ 
	m \omega k_{2} 
	+ s	\left(
		k^{2}_{2} + k^{2}_{3}
		\right)
	\right]
	\varepsilon^{\star \prime}_{\lambda^{\prime}, 2, 1}
	-
	\left(
	m \omega k_{1} + s k_{1} k_{2}
	\right)
	\varepsilon^{\star \prime}_{\lambda^{\prime}, 2, 2}
	-
	s k_{2}k_{3} 
	\varepsilon^{\star \prime}_{\lambda^{\prime}, 2, 3}
	\right]
	\right]
\nonumber
\\
& \times
	\left[
		\left(
		k \cdot p^{\prime} 
		\right) p^{\mu}
		+
		\left(
		k \cdot p
		\right) p^{\prime \, \mu}
		-
		\left(
		p \cdot p^{\prime} - M^{2}
		\right) k^{\mu}
	\right]
\nonumber
\\
& \times
	\left[
	\varepsilon^{\star \prime}_{\lambda^{\prime}, 1, \mu}
	\sin \left(\arctan\left(\frac{\Gamma_{1}}{\Gamma_{2}}\right) \right)
	\mp
	\varepsilon^{\star \prime}_{\lambda^{\prime}, 2, \mu}
	\cos \left(\arctan\left(\frac{\Gamma_{1}}{\Gamma_{2}}\right) \right)
	\right]
\nonumber
\\
% SIXTH BLOCK %%%%%%%%%%%%%%%%%%%%%%%%%%%%%%
& -
	\left[
	\left[
	- m k_{1} k_{2} 
	\varepsilon^{\star \prime}_{\lambda^{\prime}, 2, 1} 
	+	\left(
		m k_{1} k_{3} + s \omega k_{3}
		\right) 
	\varepsilon^{\star \prime}_{\lambda^{\prime}, 2, 2}
	+
	\left[
		m
		\left(
		k^{2}_{1} + k^{2}_{2}
		\right)
		+ s \omega k_{2}
	\right] 
	\varepsilon^{\star \prime}_{\lambda^{\prime}, 2, 3}
	\right]
	\right.
\nonumber
\\
&\mp
	\left.
	\left[
	\left[ 
	m \omega k_{2} 
	+ s	\left(
		k^{2}_{2} + k^{2}_{3}
		\right)
	\right]
	\varepsilon^{\star \prime}_{\lambda^{\prime}, 1, 1}
	-
	\left(
	m \omega k_{1} + s k_{1} k_{2}
	\right)
	\varepsilon^{\star \prime}_{\lambda^{\prime}, 1, 2}
	-
	s k_{2}k_{3} 
	\varepsilon^{\star \prime}_{\lambda^{\prime}, 1, 3}
	\right]
	\right]
\nonumber
\\
& \times
	\left[
		\left(
		k \cdot p^{\prime} 
		\right) p^{\mu}
		+
		\left(
		k \cdot p
		\right) p^{\prime \, \mu}
		-
		\left(
		p \cdot p^{\prime} - M^{2}
		\right) k^{\mu}
	\right]
\nonumber
\\
& \times
	\left.
	\left[
	\varepsilon^{\star \prime}_{\lambda^{\prime}, 2, \mu}
	\sin \left(\arctan\left(\frac{\Gamma_{1}}{\Gamma_{2}}\right) \right)
	\mp
	\varepsilon^{\star \prime}_{\lambda^{\prime}, 1, \mu}
	\cos \left(\arctan\left(\frac{\Gamma_{1}}{\Gamma_{2}}\right) \right)
	\right]
	\right\}
\nonumber
\\
& \times
J_{n} \left( \sqrt{\Gamma^{2}_{1} + \Gamma^{2}_{2}} \right)
J_{n \mp 1} \left( \sqrt{\Gamma^{2}_{1} + \Gamma^{2}_{2}} \right)
\, .
\label{real-part-M-M-amplitude-full-App}
\end{align}
Here, we have used the notation $\varepsilon^{\star \, \prime}_{\lambda^{\prime}, \mu} = \varepsilon^{\star \, \prime}_{\lambda^{\prime}, 1 , \mu} + i \varepsilon^{\star \, \prime}_{\lambda^{\prime}, 2 , \mu}$ for an arbitrary polarization of the emitted photon and  
$\boldsymbol{\varepsilon^{\star \prime}_{\lambda^{\prime}, 1}} = \{ \varepsilon^{\star \prime}_{\lambda^{\prime}, 1, 1}, 
\varepsilon^{\star \prime}_{\lambda^{\prime}, 1, 2}, 
\varepsilon^{\star \prime}_{\lambda^{\prime}, 1, 3} \}$ 
and
$\boldsymbol{\varepsilon^{\star \prime}_{\lambda^{\prime}, 2}} = \{ \varepsilon^{\star \prime}_{\lambda^{\prime}, 2, 1}, 
\varepsilon^{\star \prime}_{\lambda^{\prime}, 2, 2}, 
\varepsilon^{\star \prime}_{\lambda^{\prime}, 2, 3} \}$ for its three-dimensional components. Also, we have dropped the argument $k$ from the functions $\varepsilon$ for simplicity.

In concrete calculation involving the above amplitudes, it is useful to use the inverse trigonometric relations
\begin{align}
\sin \left(\arctan\left(\frac{\Gamma_{1}}{\Gamma_{2}}\right) \right)
& =
\left( \frac{\Gamma_{1}}{\Gamma_{2}} \right)
\left[1 + \left( \frac{\Gamma_{1}}{\Gamma_{2}} \right)^{2} \right]^{-1}
\, ,
\label{sin-arctan}
\\
\cos \left(\arctan\left(\frac{\Gamma_{1}}{\Gamma_{2}}\right) \right)
& =
\left[1 + \left( \frac{\Gamma_{1}}{\Gamma_{2}} \right)^{2} \right]^{-1}
\, .
\label{cos-arctan}
\end{align}

\section*{Appendix C: Properties of Bessel functions}
\renewcommand{\theequation}{C.\arabic{equation}}
\setcounter{equation}{0}

In this section, we summarize some of the properties of the Bessel functions used to calculate the partial amplitudes from Section 5. The general reference for these relations is \cite{Abramowitz:1965}.

We have used the following properties of the Bessel functions. The limits at $\zeta \to 0$ are given by 
\begin{align}
J_{n}(\zeta) & \simeq \frac{1}{\Gamma(n + 1)}
	\left(
	\frac{\zeta}{2}
	\right)^{n}
\, ,
\label{Jn-to-zero}
\\
J_{0} ( \zeta )  & \simeq 1
\, ,
\label{J0-to-zero}
\end{align}
where $\Gamma(n + 1)$ is the gamma function for natural numbers defined as
\begin{equation}
\Gamma (n) = 1 \cdot 2 \cdot 3\cdots (n-1)=(n-1)!
\, ,
\qquad
\label{gamma-function}
\Gamma (n+1)=n \Gamma (n)
\, ,
\qquad
\Gamma (1) = 1
\, .
\end{equation}
The asymptotic form at $\zeta \to \infty$ is 
\begin{equation}
J_{n }(\zeta) \simeq 
\sqrt{\frac {2}{\pi \zeta}} 
\cos\left( 
		\zeta - 
		\frac{n \pi }{2} 
		-
		\frac{\pi }{4} 
	\right)
\, .
\label{Jn-to-infty}
\end{equation}
The modified Bessel functions can be expressed in terms of Bessel functions. Also they have a symmetry property in their index. The relevant relations are
\begin{align}
I_{-n} \left( i \zeta \right) & = I_{n} \left(i \zeta \right)
\, \
\label{J-minus-plus-1}
\\
I_{n} \left( i \zeta \right) & = e^{\frac{n \pi i}{2}} J_{n} \left( \zeta \right)
\, ,
\label{J-n-n}
\\
I_{0} \left( i \zeta \right) & =
J_{0} \left( \zeta \right) 
\, .
\label{I0-J0}
\end{align}
From these properties, it follows that
\begin{equation}
\vert I_{0} \left( i \zeta \right) \vert^{2} = 
\vert J_{0} \left( \zeta \right) \vert^{2}
\, ,
\quad
\vert I_{-1} \left( i \zeta \right) \vert^{2} = 
\vert J_{-1} \left( \zeta \right) \vert^{2}
\, ,
\quad
\vert I_{+1} \left( i \zeta \right) \vert^{2} = 
\vert J_{+1} \left( \zeta \right) \vert^{2}
\, . 
\label{moduli}
\end{equation}

%--------- REFERENCES -------------------------------------

\end{document}